\documentclass[final,1p,times]{elsarticle}

\usepackage{amsmath,amssymb,bm}

\journal{Annals of Physics}

\begin{document}

\begin{frontmatter}



\title{The issue of gauge choice in the Landau problem and the
physics of canonical and mechanical orbital angular momenta}


\author[label1,label2]{M.~Wakamatsu\corref{cor1}}
\ead{wakamatu@post.kek.jp}
\author[label1]{Y.~Kitadono}
\author[label1]{P.-M.~Zhang}

\cortext[cor1]{corresponding author}

\address[label1]{Institute of Modern Physics, Chinese Academy of Sciences,\\
Lanzhou, People's Republic of China, 730000}
\address[label2]{KEK Theory Center, Institute of Particle and Nuclear Studies,\\
High Energy Accelerator
Research Organization (KEK),\\
1-1, Oho, Tsukuba, Ibaraki 305-0801, Japan}

\begin{abstract}
One intriguing issue in the nucleon spin decomposition problem is
the existence of two types of decompositions, which are representably
characterized by two different orbital angular momenta (OAMs) of
quarks. The one is the mechanical OAM,
while the other is the so-called gauge-invariant canonical (g.i.c.) OAM,
the concept of which was introduced by Chen et al. 
An especially delicate quantity is the g.i.c. OAM, which must be distinguished
from the ordinary (gauge-variant) canonical OAM. 
We find that, owing to its analytically solvable nature, the famous Landau
problem offers an ideal tool to understand the difference and
the physical meaning of the above three OAMs, i.e. the standard canonical
OAM, g.i.c. OAM, and the mechanical OAM.  
We analyze these three OAMs in two different formulations
of the Landau problem, first in the standard (gauge-fixed) formulation
and second in the gauge-invariant (but path-dependent) formulation of
DeWitt. Especially interesting is the latter formalism. It is shown that the
choice of path has an intimate connection with the choice of gauge,
but they are not necessarily equivalent. Then, we answer the
question about what is the consequence of a particular choice of path in
DeWitt's formalism. This analysis also clarifies the implication of the gauge
symmetry hidden in the concept of g.i.c. OAM.
Finally, we show that the finding above offers a clear
understanding about the uniqueness or non-uniqueness problem of
the nucleon spin decomposition, which arises from the arbitrariness in
the definition of the so-called physical component of the gauge field.
\end{abstract}


\end{frontmatter}



\section{Introduction}
\label{Section:s1}

How the total nucleon spin can be decomposed 
into the spin and orbital angular momentum (OAM) of quarks and
gluons without conflicting with the gauge-invariance principle is a
very delicate problem of quantum chromodynamics (QCD), and
a lot of debates were developed for a considerable time.
(See \cite{Review_LL14},\cite{Review_Waka14},\cite{Review_LL16}
for review.)
A consensus now is that there are two different types of
complete decomposition of the nucleon spin, which are
respectively called the canonical type decomposition and
the mechanical type one \cite{Waka2010},\cite{Waka2011}. 
Unfortunately, there still remains different opinions on the proper
physical interpretation of the two decompositions of the nucleon spin. 
Since these two decompositions are representably
characterized by the canonical OAM and the mechanical OAM
of quarks, the judgment on the merits
and demerits of the two nucleon spin decomposition cannot
be delivered unless we {\it do} understand the difference
and the physical meanings of these two OAMs correctly.

To understand the difference of these two OAMs in a
clearest fashion, the familiar Landau problem turns out to offer an
ideal testing ground, because of its analytically solvable nature.
The Landau problem is a quantum mechanical problem,
which describes the motion of an electron or any charged
particle in a uniform magnetic
field \cite{Landau1930},\cite{Landau-Lifschitz}.
Since the Hamiltonian of the Landau problem contains
the vector potential not the magnetic field itself,
it has a freedom of gauge choice. 
In the standard treatment of the Landau problem, one first
fixes a gauge and then solves the quantum mechanical
Schr\"{o}dinger equation. Typical gauge choices are the
two types of Landau gauge and the symmetric gauge.
Less popular is a gauge-invariant formulation of the
Landau problem, in which the Schr\"{o}dinger equation can
be solved without assuming any particular form of vector
potential, which means that the problem can be
solved without fixing gauge. 

The purpose of the present study is to clearly understand the
physical meaning of the two OAMs, i.e. the canonical OAM
and the mechanical OAM of the electron, by making full use
of the analytically solvable nature of the Landau problem. 
What plays an important role in this analysis is the
existence of another OAM called the pseudo angular momentum in
the literature \cite{Konstantinou2016},\cite{Konstantinou2017}.
(A similar quantity called the pseudo momentum was
already discussed in \cite{Yoshioka2002}.)
A unique feature of this quantity is that it is gauge-invariant,
and reduces to the canonical OAM in a suitable gauge. 
Because of this property, the pseudo OAM is very much
resembling the quantity called the gauge-invariant-canonical
(g.i.c.) OAM, the concept of which was introduced by Chen et al.
in the discussion of the nucleon spin decomposition problem
as well as the nucleon momentum decomposition 
problem \cite{Chen2008},\cite{Chen2009}.
(In the following, we mainly use the latter name instead of
the terminology pseudo angular momentum.)
In view of this situation, we are faced with three types of OAM.
The first is the ordinary canonical OAM, which is usually
believed to be a gauge-variant quantity. The second is the
g.i.c. OAM. The last is the manifestly gauge-invariant mechanical
OAM. We aim at clarifying the physical meaning of these three OAMs,
by paying special attention to their gauge-dependence or
gauge-independence. 
We shall carry out this analysis in both of the standard
(gauge-fixed) formulation of the Landau problem and also in
the gauge-invariant formulation of the Landau problem.
It turns out that this parallel analysis is particularly useful not
only for clarifying the physical meaning of the
above three OAMs but also for getting deeper understanding of the
delicacy hidden in the gauge choices in the Landau problem
as well as in the nucleon spin decomposition problem.

The plan of the paper is as follows. In section \ref{Section:s2},
a theoretical analysis of the above-mentioned three OAMs
is carried out within the framework of the standard
formulation of the Landau problem. We first show that the
standard eigen-functions of the Landau Hamiltonian in the Landau
gauge is not connected with the standard eigen-functions of the
symmetric gauge by a simple gauge transformation.
And then, we derive the correct relation between these two
eigen-functions in the two different gauges. 
Next in section \ref{Section:s3}, the same
three OAMs are analyzed within the framework of the gauge-invariant
formulation of the Landau problem.
The delicacy of the gauge choice in the Landau problem
will become clear by combining the analysis in this section
and that in the previous section. 
In section \ref{Section:s4}, we discuss in some detail the
similarity between the concept of pseudo momentum and
pseudo OAM and that of 
g.i.c. momentum and g.i.c. OAM of Chen et al.
This consideration is useful for unraveling the delicacy hidden
in the idea of g.i.c. momentum and g.i.c. OAM.
Next in section \ref{Section:s5}, we make some important
remarks on the issue of gauge choice in the nucleon spin decomposition
problem. 
Putting all the knowledge obtained so far in force from our general viewpoint
of the gauge-invariant but path-dependent formulation of gauge theory, 
we will clarify what is really meant by the idea of the physical component
of the gauge field, which was introduced by Chen et al.
Finally in section  \ref{Section:s6}, we summarize what we have found and
make some concluding remarks.

\section{Canonical and mechanical orbital angular momenta in the standard
formulation of the Landau problem}
\label{Section:s2}

The Landau problem is a very fundamental quantum mechanical
problem, which describes the motion of an electron, or any charged
particle, under the influence of uniform magnetic field.
In the following, the direction of the magnetic field is taken in the
$z$-direction, and the mass and the charge of the electron are denoted
as $m_e$ and $- \,e$ with $e > 0$. 
(The natural unit $c = \hbar = 1$ is used throughout the paper.) 

The Hamiltonian of the system is represented as
\begin{equation}
 H \ = \ \frac{\bm{\Pi}^2}{2 \,m_e} ,
\end{equation}
in terms of the so-called mechanical (or kinetic) momentum
$\bm{\Pi}$ defined as
\begin{equation}
 \bm{\Pi} \ = \ \bm{p} \ + \ e \, \bm{A} ,
\end{equation}
with $\bm{A}$ being the vector potential giving the magnetic field through
the relation $\nabla \times \bm{A} = \bm{B} = B \,\bm{e}_z$.

As is well-known, the choice of vector potential, which gives the same magnetic
field, is not unique. We say that there is a freedom of gauge choices.
The most popular gauge choices are the following three : 
\begin{eqnarray}
 \mbox{1st Landau gauge} \ &:& \ 
 \bm{A} \ = \ \bm{A}_{L_1} \ = \ - \,B \,y \,\bm{e}_x, \\
 \mbox{2nd Landau gauge} \ &:& \ 
 \bm{A} \ = \ \bm{A}_{L_2} \ = \ + \,B \,x \,\bm{e}_y, \\
 \mbox{symmetric gauge } \ &:& \ 
 \bm{A} \ = \ \bm{A}_{S} \ = \ - \,\frac{1}{2} \,B \,y \,\bm{e}_x
 \ + \ \frac{1}{2} \,B \,x \,\bm{e}_y. 
\end{eqnarray}
The two Landau gauges can be treated almost in the same manner, so that
in this section we consider the 2nd Landau gauge and call it simply
the Landau gauge.

Our primary concern in the present paper is to unravel the physical
meaning of the two types of orbital angular momentum (OAM), i.e. 
the canonical OAM,
\begin{equation}
 L_{can} \ \equiv \ (\bm{r} \times \bm{p})_z \ = \ 
 - \,i \,\left( x \,\frac{\partial}{\partial y} \ - \ 
 y \,\frac{\partial}{\partial x} \right),
\end{equation}
and the mechanical OAM,
\begin{equation}
 L_{mech} \ \equiv \ (\bm{r} \times \bm{\Pi})_z \ = \ (\bm{r} \times \bm{p})_z
 \ + \ e \,(\bm{r} \times \bm{A})_z .
\end{equation}
They are related as
\begin{equation}
 L_{can} \ = \ L_{mech} \ + \ L_{pot} , 
\end{equation}
where
\begin{eqnarray}
 L_{pot} \ \equiv \ - \,e \,(\bm{r} \times \bm{A})_z \ = \ 
 - \,e \,(x \,A_y \ - \ y \,A_x) . \label{Eq:gauge-dep potential OAM}
\end{eqnarray}
Here, $L_{pot}$ is basically the quantity, which was called in 
\cite{Waka2010},\cite{Waka2011} the {\it potential angular momentum}. 
(To be more rigorous, the 
potential angular momentum introduced in \cite{Waka2010},\cite{Waka2011}
is defined by
$L_{pot} = - \,e \,(\bm{r} \times \bm{A}_\perp)_z$ with $\bm{A}_\perp$
being the transverse component of $\bm{A}$, and it is gauge-invariant
under the residual gauge transformation within the Coulomb gauge.)
The potential angular momentum defined by (\ref{Eq:gauge-dep potential OAM})
is generally gauge-dependent
and takes the following form according to the choices of gauge :
\begin{eqnarray}
 \mbox{symmetric gauge} \ &:& \ 
 L_{pot} \ = \ - \,\frac{1}{2} \,e \,B \,(x^2 + y^2), \\
 \mbox{Landau gauge} \ &:& \ 
 L_{pot} \ = \ - \,e \,B \,x^2. 
\end{eqnarray}

At this stage, it is useful to recall the known solution to the Landau
problem, i.e. the eigen-values and the eigen-functions of the Landau Hamiltonian $H$.
The standardly-known eigen-functions of $H$ in the Landau gauge are
given as \cite{Landau1930},\cite{Landau-Lifschitz}
\begin{equation}
 \Psi^{(L)}_{n, k_y} (x, y) \ = \ \frac{1}{\sqrt{2 \,\pi}} \,e^{\,i \,k_y \,y} \,
 N_n \,H_n \left( \frac{x - x_0}{l_B} \right) \,
 e^{\,- \,\frac{(x - x_0)^2}{2 \,l_B^2}} ,
\end{equation}
where $H_n (x)$ are the Hermite polynomials, while the other constants
are defined by
\begin{equation}
 N_n \ = \ \left( \frac{1}{\sqrt{\pi} \,2^n \,n ! \,l_B} \right)^{1/2} ,
\end{equation}
and
\begin{equation}
 x_0 \ = \ - \,l_B^2 \,k_y \ \ \ \mbox{with} \ \ \ 
 l_B^2 \ = \ \frac{1}{e \,B} .
\end{equation}
The quantity $x_0$ is sometimes called the guiding center, while $l_B$
is called the magnetic length.
The above wave functions are the simultaneous eigen-functions of the
canonical momentum operator $p_y$ and the Landau Hamiltonian $H$ : 
\begin{eqnarray}
 p_y \,\Psi^{(L)}_{n, k_y} (x, y) &=& k_y \,\Psi^{(L)}_{n, k_y} (x, y), \\
 H \,\Psi^{(L)}_{n, k_y} (x, y) &=& \left( n + \frac{1}{2} \right) \,\omega \,
 \Psi^{(L)}_{n, k_y} (x, y) \hspace{8mm} (n = 0, 1, 2, \cdots) ,
\end{eqnarray}
where $n = 0, 1, 2, \cdots$, and $\omega = \frac{e \,B}{m_e}$.
Note that the eigen-energies depend only on the so-called Landau
quantum number $n$, and they do not depend on the quantum
number $k_y$. 

On the other hand, the standard eigen-functions in the symmetric gauge are
given as \cite{Fan-Lin2000},\cite{Li-Wang1999}
\begin{equation}
 \Psi^{(S)}_{n,m} (x,y) \ = \ \frac{1}{\sqrt{2 \,\pi}} \,e^{\,i \,m \,\phi} \,
 N_{n,m} \,\left( \frac{r^2}{2 \,l_B^2} \right)^{|m|/2} \,
 e^{\,- \,\frac{r^2}{4 \,l_B^2}} \,
 L^{|m|}_{n - \frac{|m| + m}{2}} \left( \frac{r^2}{2 \,l_B^2} \right) , \label{Eq:SG_sol}
\end{equation}
with $m$ being an integer satisfying the condition $m \leq n$,
$L_n^m (x)$ is the associated Laguerre polynomial, while
\begin{equation}
 N_{n,m} \ = \ (- \,1)^{\,n + \frac{|m| + m}{2}} \,\,\frac{1}{l_B} \,\,
 \sqrt{\frac{\left(n - \frac{|m| + m}{2}\right)}
 {\left(n + \frac{|m| - m}{2}\right)}} . \label{Eq:SG_norm}
\end{equation}
They are the simultaneous eigen-functions of the canonical OAM operator
$L_{can}$ and the Landau Hamiltonian $H$ : 
\begin{eqnarray}
 L_{can} \,\Psi^{(S)}_{n,m} (x, y) &=& m \,\Psi^{(S)}_{n,m} (x, y), \\
 H \,\Psi^{(S)}_{n,m} (x, y) &=& \left( n + \frac{1}{2} \right) \,\omega \,
 \Psi^{(S)}_{n,m} (x, y) ,
\end{eqnarray}
Here, we are interested in the expectation values of the canonical and
mechanical OAM operators in the above two gauges.
The answer to this question is well known
in the symmetric gauge. One can easily confirm that
\begin{eqnarray}
 \langle L_{can} \rangle &=& \langle \Psi^{(S)}_{n,m} (x, y) \,|\,
 L_{can} \,|\, \Psi^{(S)}_{n,m} (x, y) \rangle \ = \ m ,\\
 \langle L_{pot} \rangle &=& \langle \Psi^{(S)}_{n,m} (x, y) \,|\,
 - \,\frac{1}{2} \,e \,B \,r^2 \,|\, \Psi^{(S)}_{n,m} (x, y) \rangle 
 \ = \ - \,\left( 2 \,n \, + \, 1 \, - \, m \right) ,
\end{eqnarray}
which in turn gives
\begin{equation}
 \langle L_{mech} \rangle \ = \ \langle L_{can} \rangle \ - \ 
 \langle L_{pot} \rangle \ = \ m \ + \ (2 \,n + 1 - m) \ = \ 
 2 \,n + 1 .
\end{equation}
(We point out that the overall sign of the expectation
value of the mechanical OAM is opposite to that in \cite{Waka2016}.
This is because we consider here the motion of the
electron with negative charge in contrast to \cite{Waka2016}, in which
the motion of a particle with positive charge is studied. 
Note that the cyclotron motion of a particle
with negative charge is counterclockwise in the $x$-$y$ plane, so that
its mechanical angular momentum should be positive.)

Remarkably, the expectation value of the mechanical OAM depends only
on the Landau quantum number $n$, and it does not depends on the
eigen-value $m$ of the canonical OAM. 
We also recall that the expectation value of the mechanical OAM is just proportional
to the eigen-energies of the Landau level, or the expectation value of the
Landau Hamiltonian :
\begin{equation}
 E_n \ = \ \left( n + \frac{1}{2} \right) \,\omega \ = \ 
 \frac{e \,B}{2 \,m_e} \,
 \langle \Psi^{(S)}_{n,m} \,|\,L_{mech} \,|\,\Psi^{(S)}_{n,m} \rangle .
\end{equation}
This is consistent with the
known fact that what describes the physical cyclotron motion
is the mechanical OAM not the canonical one.

One might naturally ask the following question. What are expectation
values of the two angular momentum operators in the Landau gauge ?
Since the mechanical OAM is generally believed to be a gauge-invariant
operator, one might at least anticipate that the expectation
value of the mechanical OAM operator is independent of the gauge choice,
i.e. it is the same in both the symmetric and Landau gauges. 
The problem is more subtle than
this naive expectation. To see it, imagine that we want to evaluate the
expectation value of the canonical OAM operator in the Landau gauge, or 
more precisely the expectation value between the standard eigen-states
in the Landau gauge,
\begin{equation}
 \langle L_{can} \rangle \ \equiv \ \langle \Psi^{(L)}_{n,k_y} \,|\,
 - \,i \,\left( x \,\frac{\partial}{\partial y} \ - \ 
 y \,\frac{\partial}{\partial x} \right) \,|\, \Psi^{(L)}_{n,k_y} \rangle . 
\end{equation}
First, this expression must be taken with care, because the
plane-wave in the $y$-direction is not normalizable. To make the
expression finite, the box normalization was adopted
in \cite{Waka2016}, which means the replacement 
\begin{equation}
 \frac{1}{\sqrt{2 \,\pi}} \,e^{\,i \,k_y \,y} \ \ \rightarrow \ \ 
 \frac{1}{\sqrt{L}} \,e^{\,i \,k_y \,y} \ \ \ \mbox{with} \ \ \ 
 - \,\frac{L}{2} \leq y \leq \frac{L}{2} .
\end{equation}
With this replacement, the expectation value of the canonical OAM
operator becomes
\begin{equation}
 \langle L_{can} \rangle \ = \ - \, \frac{k_y^2}{e B} . \label{Eq:LEV_L_can}
\end{equation}
With the same box normalization, the expectation value of the potential
OAM operator is shown to be
\begin{equation}
 \langle L_{pot} \rangle \ \equiv \ \langle \Psi^{(L)}_{n,k_y} \,|\,
 - \,e \,B \,x^2 \,|\,\Psi^{(L)}_{n,k_y} \rangle
 \ = \ - \,\left\{ \left( n + \frac{1}{2} \right) \ + \ 
 \frac{k_y^2}{e B} \right\} .  \label{Eq:LEV_L_pot}
\end{equation}
(There is a factor of two mistake in Eq.(59) of \cite{Waka2016}, 
which gives $2 \,n + 1$
instead of $n + \frac{1}{2}$ in the above equation. This led to an erroneous
conclusion that the expectation value of the  mechanical OAM operator
in the Landau gauge, or more precisely in the expectation value in the
standard eigen-functions in the Landau gauge, is the same as that
in the symmetric gauge. See below for more detail.)
The expectation value of the mechanical OAM operator can easily be
obtained from the above two relations : 
\begin{eqnarray}
 \langle L_{mech} \rangle &=& \langle L_{can} \rangle \ - \ 
 \langle L_{pot} \ \rangle \ = \ 
 n + \frac{1}{2} . 
\end{eqnarray}
We find that the $k_y^2$ dependent terms in the expectation values
of the canonical OAM and potential OAM terms cancel exactly.
However, the expectation value of the mechanical OAM in
the Landau gauge
turns out to be half of the corresponding expectation value in the
symmetric gauge. At first glance, this result appears to contradict the
standard belief that the mechanical OAM operator is a gauge-invariant
quantity. However, it would be more legitimate to say that
the $k_y^2$ dependent terms, contained in both of (\ref{Eq:LEV_L_can}) and
(\ref{Eq:LEV_L_pot}), do not have clear physical meaning at least in
its present form. This is due to the fact that, for the eigen-functions in the
Landau gauge, $y$-coordinate of the electron is totally uncertain.
After all, we conclude that
the concept of orbital angular momentum does not have
well-defined physical meaning within the standard eigen-functions
in the Landau gauge. 

The above analysis shows that the issue of the gauge-invariance of the
mechanical OAM in the Landau problem should be investigated more carefully. 
It is a widely-known fact that the two gauge choices, i.e. the symmetric
gauge and the Landau gauge, are related by the gauge transformation,
$\bm{A}_{L} = \bm{A}_{S} + \nabla \chi$ with $\chi = \frac{1}{2} \,B \,x \,y$.
To understand the true meaning of this gauge transformation, we
first briefly review the consequence of gauge-invariance in quantum
mechanics.
In quantum mechanics, the gauge transformation
\begin{equation}
 \bm{A}^\prime \ = \ \bm{A} \ + \ \nabla \chi ,
\end{equation}
must be supplemented with the phase transformation of the charged particle wave function
\begin{equation}
 \psi^\prime (\bm{r}) \ = \ e^{i \,q \,\chi (\bm{r})} \,\psi (\bm{r}) ,
\end{equation}
such that the identity
\begin{equation}
 ( \hat{\bm{p}} \ - \ q \,\bm{A}^\prime (\bm{r}) ) \,\psi^\prime (\bm{r}) \ = \ 
 e^{\,i \,q \,\chi (\bm{r})} \,(\hat{\bm{p}} \ - \ q \,\bm{A} (\bm{r}) ) \,
 \psi (\bm{r}) ,
\end{equation}
holds. The standard statement is that observables of the type
\begin{equation}
 \langle \psi_f^\prime (\bm{r}) \,|\, O_1 (\bm{r}) \,|\,\psi_i^\prime (\bm{r}) 
 \rangle \ = \ 
 \langle \psi_f (\bm{r}) \,|\, O_1 (\bm{r}) \,|\,\psi_i (\bm{r}) \rangle ,
\end{equation}
and also of the type
\begin{equation}
 \langle \psi_f^\prime (\bm{r}) \,|\, O_2 (\hat{\bm{p}} - q \,\bm{A}^\prime) \,
 |\,\psi_i^\prime (\bm{r}) 
 \rangle \ = \ 
 \langle \psi_f (\bm{r}) \,|\, O_2 (\hat{\bm{p}} - q \,\bm{A} ) \,
 |\,\psi_i (\bm{r}) \rangle , 
\end{equation}
are unchanged by the gauge transformations, i.e. they are
gauge-invariant.

The coordinate operator $\hat{\bm{r}}$ is the simplest example
of the first type of observables.
The mechanical momentum $\hat{\bm{p}} - q \,\bm{A}$ as well as the
mechanical orbital angular momentum 
$\bm{r} \times (\hat{\bm{p}} - q \,\bm{A})$ are special
examples of the 2nd types of operator. (Aside from the
above few paragraphs, the quantum
operators $\hat{\bm{r}}$ and $\hat{\bm{p}}$ are simply denoted
as $\bm{r}$ and $\bm{p}$, since no confusion is expected to occur.)
From the general statement above, one might naively anticipate that the 
following relation would hold
\begin{equation}
 \langle \Psi^{(L)} \,|\, L^{(L)}_{mech} \,|\,\Psi^{(L)} \rangle \ = \ 
 \langle \Psi^{(S)} \,|\, L^{(S)}_{mech} \,|\,\Psi^{(S)} \rangle .
\end{equation}
Here, $L^{(S)}_{mech} = [\bm{r} \times (\bm{p} + e \,\bm{A}_S)]_z$ is the mechanical
OAM operator in the symmetric gauge, while $| \, \Psi^{(S)} \rangle$ are the
eigen-states of the Landau Hamiltonian in the same gauge.
On the other hand, $L^{(L)}_{mech} = [\bm{r} \times (\bm{p} + e \,\bm{A}_L) ]_z$
is the mechanical OAM operator in the Landau gauge, while $|\,\Psi^{(L)} \rangle$
are the eigen-states of the Landau Hamiltonian in the Landau gauge.
These quantities are supposed to be related by the unitary (gauge) transformation
$U \ = \ e^{\,- \,\frac{1}{2} \,i \,e \,B \,x \,y}$ as
\begin{equation}
 |\,\Psi^{(L)} \rangle \ = \ U \,|\,\Psi^{(S)} \rangle ,
\end{equation}
and
\begin{equation}
 L^{(L)}_{mech} \ = \ U \,L^{(S)}_{mech} \,U^\dagger .
\end{equation}

However, it turns out that there is an oversight in the above naive reasoning.
This is due to the well-known degeneracy of the eigen-states of the
Landau Hamiltonian \cite{Swenson1989}.
An important point is that the eigen-states of the Landau Hamiltonian
with a given Landau quantum number $n$ has infinitely many degeneracies in
both of the symmetric gauge and the Landau gauge.
Because of this degeneracy, it happens that the standard wave functions
in the Landau gauge
are not connected with those in the symmetric gauge by a simple
gauge transformation.

To understand this complexity in more concrete and clearer way, 
let us consider the operation of the gauge
transformation matrix $U$ on the eigen-states $|\,\Psi^{(S)}_{n,m} \rangle$
in the symmetric gauge, which are characterized by the two quantum
numbers, i.e. the Landau quantum number $n$ and the eigenvalue $m$ of
the canonical OAM operator.
With use of the completeness relation for the eigen-states
$|\,\Psi^{(L)}_{n,k_y} \rangle$ in the Landau gauge, we generally have
\begin{equation}
 U \,|\,\Psi^{(S)}_{n,m} \rangle \ = \ \sum_{n^\prime} \,\int \,d k_y \,
 |\,\Psi^{(L)}_{n^\prime,k_y} \rangle \,\langle \Psi^{(L)}_{n^\prime,k_y} \,|\,
 U \,|\,\Psi^{(S)}_{n,m} \rangle , \label{Eq:completeness_relation}
\end{equation}
where
\begin{equation}
 \langle \Psi^{(L)}_{n^\prime,k_y} \,|\, U \,|\,\Psi^{(S)}_{n,m} \rangle \ \equiv \ 
 \int \,dx \,d y \,\Psi^{(L)*}_{n^\prime,k_y} (x,y) \,
 e^{\,- \, i \,\frac{1}{2} \,e \,B \,x \,y} \,\Psi^{(S)}_{n,m} (x,y) .
\end{equation}
In Appendix A, we show that the above gauge transformation matrix is diagonal
in the Landau quantum number, i.e.
\begin{equation}
 \langle \Psi^{(L)}_{n^\prime,k_y} \,|\, U \,|\,\Psi^{(S)}_{n,m} \rangle \ = \ 
 \delta_{n^\prime, n} \,\langle \Psi^{(L)}_{n,k_y} \,|\, U \,|\,\Psi^{(S)}_{n,m} \rangle .
\end{equation}
Actually, this is an anticipated fact, since the gauge transformation does not
change the Landau energy, which is characterized by the quantum number $n$.

Since the above gauge transformation matrix is a function of $k_y$ or
$x_0 = - \,l_B^2 \,k_y$,
we denote it as $U_{n,m} (x_0)$, i.e.
\begin{equation}
 \langle \Psi^{(L)}_{n,k_y} \,|\, U \,|\,\Psi^{(S)}_{n,m} \rangle \ \equiv \ 
 U_{n,m} (x_0) .
\end{equation}
Thus Eq.(\ref{Eq:completeness_relation}) can be written as
\begin{equation}
 U \,|\,\Psi^{(S)}_{n,m} \rangle \ = \ \int \,d k_y \,U_{n,m} (x_0) \,
 |\,\Psi^{(L)}_{n,k_y} \rangle . \label{Eq:U-Unm}
\end{equation}
This relation shows what is generated by the action of the gauge transformation
operator $U$ on the eigen-states $|\,\Psi^{(S)}_{n,m} \rangle$ in the symmetric
gauge is a superposition of the eigen-states $|\,\Psi^{(L)}_{n,k_y} \rangle$ in the
Landau gauge with respect to the variable $k_y$ or $x_0$.
The above transformation matrix $U_{n,m} (x_0)$ corresponds to the matrix
$R_{n,m} (x_0)$ in the paper by Haugset et al.~\cite{Haugset1993}, 
although this fact is not so
transparent, since their treatment is based on a sort of gauge-invariant formulation
of the Landau problem. In our present treatment, which is based on the standard
(gauge-fixed) formulation of the Landau problem, the weight function
$U_{n,m} (x_0)$ appearing in (\ref{Eq:U-Unm}) is just the matrix element of the gauge
transformation operator $U$ sandwiched with the standard eigen-states in the
Landau and symmetric gauges.

The calculation of the weight function $U_{n,m} (x_0)$ is elementary but a
little involved, so that the detailed derivation is shown in Appendix A.
The answer is given as
\begin{equation}
 U_{n,m} (x_0) \ = \ C_{n,m} \,H_{n-m} \left( \frac{x_0}{l_B} \right) \,
 e^{\,- \,\frac{x_0^2}{2 \,l_B^2}} ,
 \label{Eq:Mat_Unm}
\end{equation}
where $H_n (x)$ is the standard Hermite polynomial, and
\begin{equation}
 C_{n,m} \ = \ \,l_B \,\left( \frac{1}{\sqrt{\pi} \,2^{n-m} \,(n-m) ! \,l_B} \right)^{1/2} .
\end{equation}
Now, inserting (\ref{Eq:Mat_Unm}) into the r.h.s. of (\ref{Eq:U-Unm}), 
we can show that the following
relation holds
\begin{equation}
 \int \,d k_y \,U_{n,m} (x_0) \,\Psi^{(L)}_{n,k_y} (x,y) \ = \ 
 e^{\,- \,i \,\frac{1}{2} \,e \,B \,x \,y} \,\Psi^{(S)}_{n,m} (x,y) .
\end{equation}
The proof of this relation is given in Appendix B.

It is convenient to define the l.h.s. of the above equation newly as $\Psi^{(L)}_{n,m} (x,y)$,
i.e.
\begin{equation}
 \Psi^{(L)}_{n,m} (x,y) \ \equiv \ \int \,d k_y \,U_{n,m} (x_0) \,\Psi^{(L)}_{n,k_y} (x,y) .
\end{equation}
Then, the relation that we have found can be expressed as
\begin{equation}
 \Psi^{(L)}_{n,m} (x,y) \ = \ e^{\,- \,i \,\frac{1}{2} \,e \,B \,x \,y} \,
 \Psi^{(S)}_{n,m} (x,y) .  \label{Eq:relation_L-S}
\end{equation}
We may now say that the l.h.s. of the above equation gives wave functions in the Landau
gauge, which is obtained from the eigen-functions in the symmetric gauge by the simple
gauge transformation with the gauge function $\chi = \frac{1}{2} \,B \,x \,y$.
However, the important point is that they are not the standard eigen-functions in the
Landau gauge but their particular superpositions.

We are now ready to evaluate the expectation values of the orbital angular momentum operators
\begin{eqnarray}
 L_{can} &=& - \,i \,\left( x \,\frac{\partial}{\partial y} \ - \ 
 y \,\frac{\partial}{\partial x} \right) , \\
 L_{pot} &=& - \,e \,B \,x^2, \\
 L_{mech} &=& L_{can} \ - \ L_{pot} ,
\end{eqnarray}
between the superposed eigenstates $\Psi^{(L)}_{n,m} (x,y)$ in the Landau gauge.
With use of the relation (\ref{Eq:relation_L-S}), 
the expectation values of the canonical OAM and the potential OAM can be easily
obtained as
\begin{equation}
 \langle \Psi^{(L)}_{n,m} \,| \,L_{can} \,|\,\Psi^{(L)}_{n,m} \rangle
 \ = \ \langle \Psi^{(L)}_{n,m} \,| \,- \,i \,\left( x \,\frac{\partial}{\partial y} \ - \ 
 y \,\frac{\partial}{\partial y} \right) \,|\,\Psi^{(L)}_{n,m} \rangle \nonumber
 \ = \ m ,
\end{equation}
and
\begin{equation}
 \langle \Psi^{(L)}_{n,m} \,|\,L_{pot} \,|\,\Psi^{(L)}_{n,m} \rangle \ = \ 
 \langle \Psi^{(L)}_{n,m} \,|\,- \,e \,B \,x^2 \,|\,\Psi^{(L)}_{n,m} \rangle 
 \ = \ - \,( 2 \,n + 1 - m) .
\end{equation}
Combining these, we finally get
\begin{eqnarray}
 \langle \Psi^{(L)}_{n,m} \,|\,L_{mech} \,|\,\Psi^{(L)}_{n,m} \rangle
 &=&  \langle \Psi^{(L)}_{n,m} \,|\,L_{can} \,|\,\Psi^{(L)}_{n,m} \rangle
 \ - \ 
 \langle \Psi^{(L)}_{n,m} \,|\,L_{pot} \,|\,\Psi^{(L)}_{n,m} \rangle
 \nonumber \\
 &=& m \ + \ (2 \,n + 1 - m) \ = \ (2 \,n + 1) .
\end{eqnarray}
This precisely coincides with the expectation value of the mechanical
OAM operator evaluated in the symmetric gauge.
We have thus confirmed that the mechanical OAM is in fact a
gauge-invariant quantity.

\section{Canonical and mechanical orbital angular momenta in the
gauge-invariant formulation of the Landau problem}
\label{Section:s3}

First, we briefly review the general framework of 
the gauge-invariant formulation of quantum
electrodynamics (QED) {\it a la} DeWitt \cite{DeWitt1962} and 
Mandelstam \cite{Mandelstam1962}. (See also \cite{Dirac1955}.)
According to DeWitt, once an appropriate set of electron and photon fields
$(\psi (x), \,A_\mu (x))$ is given, the gauge-invariant set of
the electron and photon fields $(\tilde{\psi} (x), \,\tilde{A}_\mu (x)$) can
be constructed as
\begin{eqnarray}
 \tilde{\psi} (x) \ &\equiv& \ \ 
 e^{\,i \,e \,\Lambda (x)} \,\,\psi (x), \label{GI_electron} \\
 \tilde{A}_\mu (x) \ &\equiv& \ 
 A_\mu (x) \ - \ \partial_\mu \Lambda (x) , \label{GI_photon}
\end{eqnarray}
by introducing the function
\begin{equation}
 \Lambda (x) \ = \ \int_0^1 \,A_\sigma (z) \,\,
 \frac{\partial z^\sigma}{\partial \xi} \,\,d \xi, \label{gauge_tr_func}
\end{equation}
where $z^\mu (x,\xi)$ represents a point on the line connecting an appropriate
starting point $x_0$ (the reference point) and the point $x$ in
the 4-dimensional Minkowski space, with $\xi$ being
a parameter specifying the path with the following boundary conditions : 
\begin{eqnarray}
 &\,& z^\mu (x,1) \ = \ x^\mu, \hspace{10mm} \mbox{and} \hspace{8mm} 
 z^\mu (x, 0) \ = \ x^\mu_0, \\
 &\,& \left. \frac{\partial z^\mu}{\partial x^\lambda} \right|_{\xi = 1} 
 \ = \ \delta^\mu_\lambda, \hspace{9mm} \mbox{and} \hspace{8mm} 
 \left. \frac{\partial z^\mu}{\partial x^\lambda} \right|_{\xi = 0}
 \ = \ 0 .
\end{eqnarray}
Under an arbitrary gauge transformation for the electron and photon fields
given by
\begin{eqnarray}
 \psi (x) \ &\rightarrow& \  
 e^{\,- \,i \,e \,[\omega (x) - \omega(x_0)]} \,\psi (x), \label{Eq:Gtrans_electron} \\
 A_\mu (x) \ &\rightarrow& \  
 A_\mu (x) \ + \ \partial_\mu [\omega (x) - \omega(x_0)] \ = \ 
 A_\mu (x) \ + \ \partial_\mu \omega (x),
\end{eqnarray}
the function $\Lambda (x)$ transforms as
\begin{eqnarray}
 \Lambda (x) \ &\rightarrow& \ \int_0^1 \,
 \left( A_\sigma (z) \ + \ \partial_\sigma \omega (z) \right) \,
 \frac{\partial z^\sigma}{\partial \xi} \,d \xi \nonumber \\
 &=& \ \int_0^1 \,A_\sigma (z) \,
 \frac{\partial z^\sigma}{\partial \xi} \,d \xi \ + \
 \int_0^1 \,\frac{\partial \omega (z)}{\partial \xi} \,d \xi 
 \ = \ \Lambda (x) \ + \ [\omega (x) - \omega (x_0)]. 
\end{eqnarray}
One may notice that our gauge transformation rule (\ref{Eq:Gtrans_electron})
is somewhat unusual in the sense that it depends on the reference point
$x_0$ in addition to
the point $x$ where the field is evaluated. In the original formulation
of DeWitt, this peculiarity is hidden in some sense, since the reference
point is taken to be space-time infinity. Our formulation below needs
to take this reference point to be some point in a finite region.
As a consequence, the gauge trans,formation rule (\ref{Eq:Gtrans_electron})
is required for that the gauge-invariant electron and photon fields defined
by (\ref{GI_electron}) and (\ref{GI_photon}) are in fact
gauge-invariant, as explicitly demonstrated below.
Formally, the transformation rule (\ref{Eq:Gtrans_electron}) amounts to
combining a gauge transformation with a global phase transformation.
Especially, it just reduces to an identity at the reference point $x_0$. 
As we shall see, the above gauge transformation rule does not cause
any practical problem in our following analysis.

In any case, under the above gauge transformation, the field $\tilde{\psi} (x)$ transforms as
\begin{eqnarray}
 \tilde{\psi} (x) \ &\rightarrow& \ e^{\,i \,e \,[\Lambda (x) + (\omega (x) - \omega (x_0))]} \,\,
 e^{\,- \,i \,e \,(\omega (x) - \omega (x_0))} \,\,\psi (x) \nonumber \\
 &=& \ e^{\,i \,e \,\Lambda (x)} \,\,\psi (x) \ = \ 
 \tilde{\psi} (x),
\end{eqnarray}
i.e., the new electron wave function $\tilde{\psi} (x)$ is gauge-invariant.
The gauge-invariance of $\tilde{A}_\mu (x)$ can also be readily verified.
For the sake of completeness, we reproduce the proof.
The proof goes as follows :
\begin{eqnarray}
 \tilde{A}_\mu (x) \ &=& \ A_\mu (x) \ - \ \partial_\mu \Lambda (x) \nonumber \\
 &=& \ A_\mu \ - \ \partial_\mu \,\int_0^1\,
 A_\sigma (z) \,\frac{\partial z^\sigma}{\partial \xi} \,d \xi \nonumber \\
 &=& \ A_\mu \ - \ \int_0^1\,\partial_\nu A_\sigma (z) \,
 \frac{\partial z^\nu}{\partial x^\mu} \,\frac{\partial z^\sigma}{\partial \xi}
 \,d \xi \ - \ 
 \int_0^1\,A_\sigma (z) \,\frac{\partial}{\partial \xi} \,
 \left( \frac{\partial z^\sigma}{\partial x^\mu} \right) \,d \xi \nonumber \\
 &=& \ A_\mu \ - \ \int_0^1\,\partial_\nu A_\sigma (z) \,
 \frac{\partial z^\nu}{\partial x^\mu} \,
 \frac{\partial z^\sigma}{\partial \xi} \,d \xi \nonumber \\
 &\,& \hspace{8mm} + \   
 \int_0^1\,\partial_\nu \,A_\sigma (z) \,
 \frac{\partial z^\nu}{\partial \xi} \,\frac{\partial z^\sigma}{\partial x^\mu}
 \,d \xi \ - \ A_\sigma (z) \,
 \left. \frac{\partial z^\sigma}{\partial x^\mu} \right|^{\xi = 1}_{\xi = 0}
 \nonumber \\ 
 &=& \ A_\mu \ - \ \int_0^1\,\partial_\nu A_\sigma (z) \,
 \frac{\partial z^\nu}{\partial x^\mu} \,
 \frac{\partial z^\sigma}{\partial \xi} \,d \xi \nonumber \\
 &\,& \hspace{8mm} + \   
 \int_0^1\,\partial_\nu \,A_\sigma (z) \,
 \frac{\partial z^\nu}{\partial \xi} \,\frac{\partial z^\sigma}{\partial x^\mu}
 \,d \xi \ - \ A_\sigma (x) \,\delta_\mu{}^\sigma \nonumber \\
 &=& \ - \,\int_0^1\,
 (\,\partial_\nu \,A_\sigma \ - \ \partial_\sigma \,A_\nu \,) \,
 \frac{\partial z^\nu}{\partial x^\mu} \,
 \frac{\partial z^\sigma}{\partial \xi} \,d \xi .
\end{eqnarray}
We thus find the key expression
\begin{eqnarray}
 \tilde{A}_\mu (x) \ &=& \ 
 - \,\int_0^1 \,F_{\nu \sigma} (z) \,\,
 \frac{\partial z^\nu}{\partial x^\mu} \,
 \frac{\partial z^\sigma}{\partial \xi} \,d \xi. \label{Eq:GI_potential}
\end{eqnarray}
Since the r.h.s. of the above relation is expressed only in terms of gauge-invariant
field-strength tensor, the gauge-invariance of $\tilde{A}_\mu (x)$ is 
self-evident.
This is the essence of the gauge-invariant formulation of QED by DeWitt.
Here is a catch, however. Although the r.h.s. of (\ref{Eq:GI_potential})
is certainly gauge-invariant, it generally depends on the path connecting
the reference point $x_0$ and the point $x$. (This observation is very important,
because there are in principle infinitely many paths connecting the two
space-time points $x_0$ and $x$.)

We are now ready to apply the gauge-invariant (but path-dependent) formulation
of DeWitt to the Landau problem. Since the Landau problem is a non-relativistic
and stationary quantum mechanical problem in two spatial dimension, we can
introduce the gauge-invariant electron wave function $\tilde{\Psi} (x,y)$ by
\begin{equation}
 \tilde{\Psi} (x,y) \ \equiv \ \tilde{\Psi}^{(C)} (x, y) \ = \ 
 e^{\,i \,e \,\Lambda (x,y)} \,\Psi (x,y) ,
\end{equation}
with the path-dependent phase factor,
\begin{equation}
 \Lambda (x,y) \ = \ \int_C \,\bm{A} (\bm{x}) \cdot d \bm{x} ,
\end{equation}
where $C$ is a path connecting some reference point $(x_0, y_0)$
and the point $(x, y)$ in the two-dimensional plane. 
(We recall that the electron charge is
$- \,e$ with $e > 0$ in our notation).
Alternatively, we can express the original electron wave function
$\Psi (x,y)$ in terms of the gauge-invariant electron wave
function $\tilde{\Psi}^{(C)} (x,y)$ as
\begin{equation}
 \Psi (x,y) \ = \ U (C) \,\tilde{\Psi}^{(C)} (x, y) ,
\end{equation}
with
\begin{equation}
 U (C) \ = \ e^{\,- \,i \,e \,\int_C \,\bm{A} (\bm{x}) \cdot d \bm{x}} .
\end{equation}
In the following, two different types of path choice will be
considered, separately in two subsections.

\subsection{polygonal line paths in rectangular coordinate}

\begin{figure}[ht]
\begin{center}
\includegraphics[width=6cm]{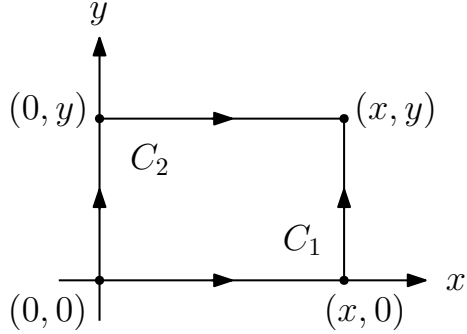}
\caption{The two polygonal line paths $C_1$ and $C_2$ defined in
the rectangular coordinate system.}
\label{Fig:line_path}
\end{center}
\end{figure}

Two simple choices of path connecting $(x_0, y_0) = (0,0)$ and
$(x, y)$ in the $x$-$y$ plane are illustrated in Fig.\ref{Fig:line_path}. 
They are made up of line segments
parallel to either of the two rectangular coordinate axes :   
\begin{eqnarray}
 C_1 \ &:& \ (0, 0) \ \rightarrow \ (x, 0) \ \rightarrow \ (x, y) ,\\
 C_2 \ &:& \ (0, 0) \ \rightarrow \ (0, y) \ \rightarrow \ (x, y) .
\end{eqnarray}
%

Selecting the path $C_1$, the relation between the standard electron
wave function $\Psi (x,y)$ and the gauge-invariant (but path-dependent)
electron wave function $\tilde{\Psi}^{(C_1)} (x,y)$ are given as
\begin{equation}
 \Psi (x,y) \ = \ U_1 \,\tilde{\Psi}^{(C_1)} (x, y) , \label{Eq:C_1_form}
\end{equation}
with
\begin{equation}
 U_1 \ \equiv \ 
 e^{\,- \,i \,e \,\int_{C_1} \,\bm{A} (\bm{x}) \cdot d \bm{x}} \ = \ 
 e^{\,- \,i \,e \,\left\{ \int_0^x \,A_x (x^\prime, 0) \,d x^\prime \ + \ 
 \int_0^y \,A_y (x,y^\prime) \,d y^\prime \right\}} .  \label{Eq:U_1} 
\end{equation}
On the other hand, with choice of the path $C_2$, we have
\begin{equation}
 \Psi (x,y) \ = \ U_2 \,\tilde{\Psi}^{(C_2)} (x, y) ,\label{Eq:C_2_form}
\end{equation}
with
\begin{equation}
 U_2 \ \equiv \  
 e^{\,- \,i \,e \,\int_{C_2} \,\bm{A} (\bm{x}) \cdot d \bm{x}} \ = \ 
 e^{\,- \,i \,e \,\left\{ \int_0^y \,A_y (0, y^\prime) \,d y^\prime \ + \ 
 \int_0^x \,A_y (x^\prime, y) \,d x^\prime \right\}} .  \label{Eq:U_2}
\end{equation}
There is a nontrivial relation between the two operators $U_1$ and $U_2$.
In fact, we have
\begin{eqnarray}
 U_1 \,U_2^{- \,1} &=& 
 e^{\,- \,i \,e \,\int_{C_1} \,\bm{A} (\bm{x}) \cdot d \bm{x}} \,
 e^{\,+ \,i \,e \,\int_{C_2} \,\bm{A} (\bm{x}) \cdot d \bm{x}} \nonumber \\
 &=& e^{\,- \,i \,e \,\oint_{C_1 - C_2} \,\bm{A} (\bm{x}) \cdot d \bm{x}} \ = \ 
 e^{\,- \,i \,e \,\iint_S \,(\nabla \times \bm{A} (\bm{x})) \cdot d \bm{S}} .
\end{eqnarray}
With use of the Stokes theorem, this gives
\begin{equation}
 U_1 \,U_2^{- \,1} \ = \ e^{\,- \,i \,\,e \,B \,x \,y}. \label{Eq:U_1_U_2_relation}
\end{equation}

To get some insight into the implication of the two path choices, we find it useful
to examine the gauge-invariant photon fields corresponding to the two paths
$C_1$ and $C_2$. First, for the path choice $C_1$, the gauge-invariant photon
fields defined by Eq.(\ref{Eq:GI_potential}) becomes
\begin{equation}
 \tilde{\bm{A}} (x,y) \ \equiv \ \tilde{\bm{A}}^{(C_1)} (x,y) \ = \ 
 - \,B \,y \,\bm{e}_x , \label{Eq:Atilde_C_1}
\end{equation}
which is nothing but the gauge potential in the 1st Landau gauge.
On the other hand, for the choice of path $C_2$, the
gauge-invariant photon field defined by Eq.(\ref{Eq:GI_potential}) 
reduces to
\begin{equation}
 \tilde{\bm{A}} (x,y) \ \equiv \ \tilde{\bm{A}}^{(C_2)} (x,y) \ = \ 
 B \, x \,\bm{e}_y ,  \label{Eq:Atilde_C_2}
\end{equation}
which just coincides with the gauge potential in the 2nd Landau gauge.

In the following, we adopt the 1st form of representation (\ref{Eq:C_1_form})
corresponding to the choice of path $C_1$. 
We have already pointed out that the gauge-invariant
photon field corresponding to this path takes the form (\ref{Eq:Atilde_C_1}).
This indicates that, in this setting of the gauge-invariant formulation,
the Landau problem would reduce to that in the 1st Landau gauge.
In the following, we shall show that it is indeed the case. To confirm it,
we start with the original Schr{\"o}dinger equation
\begin{equation}
 H \,\Psi (x,y) \ = \ E \,\Psi (x,y),
\end{equation}
where $H$ is the standard Landau Hamiltonian given as
\begin{equation}
 H \ = \ \frac{\bm{p}^2}{2 \,m_e} \ + \ \frac{e}{m_e} \,\bm{A} \cdot \bm{p}
 \ - \ i \,\frac{e}{2 \,m_e} \,(\nabla \cdot \bm{A}) \ + \ 
 \frac{e^2}{2 \,m_e} \,\bm{A}^2 .
\end{equation}
Using the relation (\ref{Eq:C_1_form}), we can transform the above eigen-equation
into the equation for the gauge-invariant electron wave functions 
$\tilde{\Psi}^{(C_1)} (x,y)$ as 
\begin{equation}
 \tilde{H}_1 \, \tilde{\Psi}^{(C_1)} (x,y) \ = \ E \,\tilde{\Psi}^{(C_1)} (x,y) ,
 \label{Eq:H_1_tilde_eq}
\end{equation}
where
\begin{equation}
 \tilde{H}_1 \ \equiv \ U_1^{- \,1} \,H \, U_1 .
\end{equation}
After some elementary algebra, we find that
\begin{equation}
 \tilde{H}_1 \ = \ 
 \frac{\bm{p}^2}{2 \,m_e} \ + \ \frac{e}{m_e} \,\tilde{\bm{A}}^{(C_1)} \cdot \bm{p}
 \ - \ i \,\frac{e}{2 \,m_e} \,(\nabla \cdot \tilde{\bm{A}}^{(C_1)}) \ + \ 
 \frac{e^2}{2 \,m_e} \,\left(\tilde{\bm{A}}^{(C_1)} \right)^2 ,
\end{equation}
where $\tilde{\bm{A}}^{(C_1)}$ is given by (\ref{Eq:Atilde_C_1}).
This means that the transformed Hamiltonian $\tilde{H}_1$ formally coincides
with the Landau Hamiltonian in the 1st Landau gauge. Explicitly, it takes
the form,
\begin{equation}
 \tilde{H}_1 \ = \ \frac{1}{2 \,m_e} \,p_x^2 \ + \ \frac{1}{2 \,m_e} \,p_y^2 \ + \ 
 \frac{e^2 \,B^2}{2 \,m_e} \,y^2 \ - \ \frac{e \,B}{m_e} \,y \,p_x .
\end{equation}
As is well-known, since this Hamiltonian does not contain $x$-dependent potential
term, $\tilde{\Psi}^{(C_1)} (x,y)$ has the following form of solutions :
\begin{equation}
 \tilde{\Psi}^{(C_1)} (x,y) \ = \ e^{\,i \,k_x \,x} \,Y (y) , \label{Eq:C_1_separable}
\end{equation}
where $k_x$ is the eigenvalue of the canonical momentum operator
$p_x = - \,i \,\frac{\partial}{\partial x}$ as
\begin{equation}
 p_x \,\tilde{\Psi}^{(C_1)} (x,y) \ = \ k_x \,\tilde{\Psi}^{(C_1)} (x,y) .
 \label{Eq:eigen_eq_p_x}
\end{equation}
Putting (\ref{Eq:C_1_separable}) into (\ref{Eq:H_1_tilde_eq}), one finds that
$Y (y)$ satisfies the following equation :
\begin{equation}
 \left[ - \,\frac{1}{2 \,m_e} \,\frac{d^2}{d y^2} \ + \ 
 \frac{1}{2 \,m_e} \,(k_x - e \,B \,y)^2 \right] \,Y (y) \ = \ 
 E \,Y (y) .
\end{equation}
This is nothing but an equation for a one-dimensional harmonic oscillator
with shifted center of oscillation, whose solution is well-known. 
They are given by
\begin{equation}
 \tilde{\Psi}_{n,k_x}^{(C_1)} (x,y) \ = \ e^{\,i \,k_x \,x} \,Y_n (y) , 
\end{equation}
with
\begin{eqnarray}
 Y_n (y) &=& \frac{1}{\sqrt{2 \,\pi}} \,N_n \,H_n \left( \frac{y - y_0}{l_B} \right) \,
 e^{\,- \,\frac{(y - y_0)^2}{2 \,l_B^2}} . \label{Eq:Landau1_Y}
\end{eqnarray}
Then, the original electron wave functions $\Psi (x,y)$ are given by
\begin{equation}
 \Psi (x,y) \ \equiv \ \Psi^{(C_1)}_{n,k_x} (x,y) \ = \ U_1 \,\tilde{\Psi}^{(C_1)}_{n,k_x} (x,y).
\end{equation}
Since the form of the original electron wave functions obtained in the above way 
depend on the chosen path $C_1$ by construction, we have explicitly written
them as $\Psi^{(C_1)}_{n, k_x} (x,y)$.

Incidentally, the eigen-equation (\ref{Eq:eigen_eq_p_x}) for the canonical
momentum can also be transformed back to an equation for the
original electron wave functions. It reads as
\begin{equation}
 U_1 \,p_x \,U_1^{- \,1} \,\Psi^{(C_1)}_{n,k_x} (x,y) \ = \ k_x \,
 \Psi^{(C_1)}_{n,k_x} (x,y) .
\end{equation}
A simple manipulation shows that
\begin{equation}
 U_1 \,p_x \,U_1^{- \,1} \ = \ p_x \ + \ e \,A_x \ + \ e \,B \,y
 \ = \ K_x  .
\end{equation}
Note that the r.h.s. of this equation is nothing but the $x$-component of
the pseudo-momentum operator $\bm{K}$ intensively discussed  
by Konstantinou and Moulopoulus \cite{Konstantinou2016},\cite{Konstantinou2017} :
\begin{equation}
 \bm{K} \ = \ \bm{p} \ + \ e \,\bm{A} \ + \ e \,\bm{r} \times \bm{B} 
 \ = \ \bm{\Pi} \ + \ e \,\bm{r} \times \bm{B}.
\end{equation}

Summarizing the analysis so far, the electron wave functions obtained in the
above way are the simultaneous eigen-states of the operator
$K_x$ and the Landau Hamiltonian $H$ : 
\begin{eqnarray}
 K_x \,\Psi_{n,k_x}^{(C_1)} (x,y) &=& k_x \,\Psi_{n,k_x}^{(C_1)} (x,y), \\
 H \,\Psi_{n,k_x}^{(C_1)} (x,y) &=& E_n \,\Psi_{n,k_x}^{(C_1)} (x,y) \ = \ 
 \left( n + \frac{1}{2} \right) \,\omega \,\Psi_{n,k_x}^{(C_1)} (x,y) ,
\end{eqnarray}
where $\omega = \frac{e \,B}{m_e}$.  Furthermore, if
we use the relation (\ref{Eq:U_1_U_2_relation}), the eigen-functions 
$\Psi^{(C_1)}_{n,k_y} (x,y)$ can be written in the following form :  
\begin{eqnarray}
 \Psi^{(C_1)}_{n,k_x} (x,y) &=& U_1 \,\tilde{\Psi}^{(C_1)}_{n,k_x} (x,y) 
 \ = \ U_2 \,e^{\,- \,i \,e \,B \,x \,y} \,\tilde{\Psi}^{(C_1)}_{n,k_x} (x,y) \nonumber \\
 &=& e^{\,i \,k_x \,x} \,e^{\,- \,i \,e \,B \,x \,y} \,
 e^{\,- \,i \,e \,\left\{ \int_0^y \,A_y (0,y^\prime) \ + \ 
 \int_0^x \,A_x (x^\prime, y) \,d x^\prime \right\}} \,Y_n (y) . \label{Eq:Landau1_sol}
\end{eqnarray}
This precisely coincides with the expression given in the paper by 
Konstantinou and Moulopoulus by a totally different flow of the logic.
As emphasized by them, a remarkable fact is that 
Eq. (\ref{Eq:Landau1_sol}) combined with
(\ref{Eq:Landau1_Y}) provide us with the eigen-functions
of the Landau Hamiltonian for arbitrary form of the vector potential $\bm{A}$.
However, one should not forget about the fact that these solutions
contain path-dependent (nonlocal) phase factor, sometimes called the Wilson line. 

The case, in which the path $C_2$ is chosen, can be treated in
a similar manner, so that we do not repeat it.
After all, the electron wave functions $\Psi^{(C_2)} (x,y)$ obtained with this
choice of path $C_2$ are the simultaneous eigen-states of the 
pseudo-momentum operator $K_y$
and the Landau Hamiltonian $H$ :
\begin{eqnarray}
 K_y \,\Psi_{n,k_y}^{(C_2)} (x,y) &=& k_y \,\Psi_{n,k_y}^{(C_2)} (x,y), \\
 H \,\Psi_{n,k_y}^{(C_2)} (x,y) &=& E_n \,\Psi_{n,k_y}^{(C_2)} (x,y) \ = \ 
 \left( n + \frac{1}{2} \right) \,\omega \,\Psi_{n,k_y}^{(C_2)} (x,y) ,
\end{eqnarray}
where
\begin{equation}
 K_y \ \equiv \ p_y \ + \ e \,A_y \ - e \,B \,x .
\end{equation}
They are explicitly expressed as
\begin{equation}
 \Psi_{n,k_y}^{(C_2)} (x,y) \ = \ U_2 \,e^{\,i \,k_y \,y} \,X_n (x) ,
\end{equation}
with
\begin{equation}
 U_2 \ = \ U_1 \, e^{\,i \,e \,B \,x \,y} \ = \ 
 e^{\,i \,e \,B \,x \,y} \,
 e^{\,- \,i \,e \,\left\{ \int_0^x \,A_x (x^\prime, 0) \,d x^\prime \ + \ 
 \int_0^y \,A_y (x , y^\prime) \,d y^\prime \right\}} , 
\end{equation}
and
\begin{equation}
 X_n (x) \ = \ \frac{1}{\sqrt{2 \,\pi}} \,N_n \,H_n \left( \frac{x - x_0}{l_B} \right) \,
 e^{\,- \,\frac{(x - x_0)^2}{2 \,l_B^2}} .
\end{equation}

Now we have two types of eigen-functions of the Landau Hamiltonian in
the rectangular coordinate. The one is the simultaneous eigen-states of the
pseudo momentum $K_x$ and the Landau Hamiltonian $H$.
The other is the simultaneous eigen-states of the pseudo momentum
$K_y$ and $H$. To understand the physical meaning of these
eigen-functions as well as the meaning of gauge choice in the Landau problem,
it is useful to investigate the function of three different types
of momentum operators, i.e. the canonical momentum $\bm{K}^{can}$, 
the mechanical momentum $\bm{K}^{mech}$, and the pseudo momentum
$\bm{K}$. They are given by
\begin{equation}
 K^{can}_x \ \equiv \ p_x, \ \ \ 
 K^{mech}_x \ \equiv \ p_x \ + \ e \,A_x, \ \ \ 
 K_x \ \equiv \ p_x \ + \ e \,A_x \ + \ e \,B \,y ,
\end{equation}
and
\begin{equation}
 K^{can}_y \ \equiv \ p_y, \ \ \ 
 K^{mech}_y \ \equiv \ p_y \ + \ e \,A_y, \ \ \ 
 K_y \ \equiv \ p_y \ + \ e \,A_y \ - \ e \,B \,x .
\end{equation}

Summarized below is the effect of operation of the three
types of operators,
$K^{can}_x, K^{mech}_x$, and $K_x$ on the states $|\,\Psi^{(C_1)}_{n,k_x} \rangle$,
which are simultaneous eigenstates of the operator $K_x$ and $H$ :

\vspace{3mm}
\noindent
1) 1st Landau gauge \ : \ $\bm{A}_{L_1} = - \,B \,y \,\bm{e}_x$
\begin{eqnarray}
 K_x^{can} \,\Psi_{n,k_x}^{(C_1)} (x,y) &=& k_x \,\Psi_{n,k_x}^{(C_1)} (x,y) , \\
 K_x^{mech} \,\Psi_{n,k_x}^{(C_1)} (x,y) &=& (k_x - e \,B \,y) \,\Psi_{n,k_x}^{(C_1)} (x,y), \\
 K_x \,\Psi_{n,k_x}^{(C_1)} (x,y) &=& k_x \,\Psi_{n,k_x}^{(C_1)} (x,y).
\end{eqnarray}

\noindent
2) 2nd Landau gauge \ : \ $\bm{A}_{L_2} = B \,x \,\bm{e}_y$
\begin{eqnarray}
 K_x^{can} \,\Psi_{n,k_x}^{(C_1)} (x,y) &=& (k_x - e \,B \,y) \,\Psi_{n,k_x}^{(C_1)} (x,y), \\
 K_x^{mech} \,\Psi_{n,k_x}^{(C_1)} (x,y) &=& (k_x - e \,B \,y) \,\Psi_{n,k_x}^{(C_1)} (x,y), \\
 K_x \,\Psi_{n,k_x}^{(C_1)} (x,y) &=& k_x \,\Psi_{n,k_x}^{(C_1)} (x,y).
\end{eqnarray}

\noindent
3) symmetric gauge \ : \ $\bm{A}_{S} = - \,\frac{1}{2} \,B \,y \,
\bm{e}_x + \frac{1}{2} \,B \,x \,\bm{e}_y$
\begin{eqnarray}
 K_x^{can} \,\Psi_{n,k_x}^{(C_1)} (x,y) &=& (k_x - \frac{1}{2} \,e \,B \,y) \,\Psi_{n,k_x}^{(C_1)} (x,y), \\
 K_x^{mech} \,\Psi_{n,k_x}^{(C_1)} (x,y) &=& (k_x - e \,B \,y) \,\Psi_{n,k_x}^{(C_1)} (x,y), \\
 K_x \,\Psi_{n,k_x}^{(C_1)} (x,y) &=& k_x \,\Psi_{n,k_x}^{(C_1)} (x,y).
\end{eqnarray}

One can see that the results of operation
of the canonical momentum operator $K^{can}_x$ are all
different depending on three gauge choices. 
This is thought to be a manifestation of gauge-variant nature
of the canonical momentum. On the other hand, the result of operation of the
operators $K^{mech}_x$ and $K_x$ are all independent of the gauge choice.
It is only natural, because both of $K^{mech}_x$ and $K_x$ are
manifestly gauge-invariant operators. 
Note, however, that the operations of $K^{mech}_x$ and $K_x$ apparently give
different answers. This last point may be seen in more obvious way through the
comparison of the expectation values.


\begin{table}[htb]
\caption{The expectation values of the three types of momentum operators,
$K^{can}_x, K^{mech}_x$, and $K_x$ on the states $|\,\Psi^{(C_1)}_{n,k_x} \rangle$.}
\label{Table:EV_Kx_eigenstate}
\vspace{4mm}
\renewcommand{\arraystretch}{1.2}
\begin{center}
\begin{tabular}{cccc}
 \hline 
 \ & \  
 $\bm{A}_{L_1} = - \,B \,y \,\bm{e}_x$ \ \ &
 $\bm{A}_{L_2} = B \, x \,\bm{e}_y$ \ \ &
 $\bm{A}_S = - \,\frac{1}{2} \,B \, y \,\bm{e}_x + \frac{1}{2} \,B \,x \,\bm{e}_y$ \ \\
 \hline
 $\langle K_x^{can} \rangle$ & 
 \ $k_x$ \ &
 \ $0$ \ & 
 \ $\frac{1}{2} \,k_x$ \ \\
 \hline
 $\langle K_x^{mech} \rangle$  &  
 \ $0$ \ &
 \ $0$ \ & 
 \ $0$ \ \\
 \hline
 $\langle K_x \rangle$ & 
 \ $k_x$ \ &
 \ $k_x$ \ & 
 \ $k_x$ \ \\
 \hline
\end{tabular}
\end{center}
\end{table}

Table \ref{Table:EV_Kx_eigenstate} shows the comparison of
the expectation values of the three operators $K^{can}_x$, $K^{mech}_x$, and
$K_x$ in the same eigenstates $| \,\Psi^{(C_1)}_{n, k_x} \rangle$ with different
gauge choices.
One confirms that the expectation values of the canonical momentum
operator $K^{can}_x$ is gauge-dependent. On the other hand, the expectation
values of the mechanical momentum operator $K^{mech}_x$ are independent of
the gauge choices. This is again an expected result for the mechanical
momentum, since it is a gauge-invariant quantity. What is less familiar is the role
of another gauge-invariant momentum, which is called the pseudo momentum
$K_x$ in the 
literatures \cite{Konstantinou2016},\cite{Konstantinou2017},\cite{Yoshioka2002}. 
The expectation values of this operator
take the same value $k_x$ independently of the choices of gauge. Furthermore,
it coincides with the expectation value of the canonical momentum
operator $K_x$ in the 1st Landau gauge.
This is not surprising, if we remember the following two facts.
First, $|\,\Psi^{(C_1)}_{n,k_x} \rangle$ are the
eigen-functions of the pseudo-momentum operator $K_x$ with the
eigenvalue $k_x$ with any choices of gauge.
Second, the pseudo-momentum operator $K_x$ reduces to the
canonical momentum operator $K^{can}_x$ in the 1st Landau gauge. 
However, one should clearly recognize the fact that an expectation value $k_x$
of the operator $K_x$ is apparently different from
that of the mechanical momentum operator $K^{mech}_x$, which is {\it zero}.
As such, although $K^{mech}_x$ and $K_x$ are both gauge-invariant, they have
totally different physical meaning. 

To understand the importance of this difference,
we first recall that an electron in a uniform magnetic field makes an circular
motion around some center. 
Obviously, the $x$-component of the electron's momentum averaged over the
period of this circular motion is zero. The fact that the expectation values of
the mechanical momentum operator $K^{mech}_x$ are zero irrespectively of the
gauge choices is completely consistent with this observation, which is not
unrelated to the physical nature of the mechanical momentum.
On the other hand, we have seen that the expectation values of the
pseudo-momentum operator $K_x$ are $k_x$, which are generally nonzero.
(We recall once again that the states $|\,\Psi^{(C_1)}_{n, k_x} \rangle$ are the
eigen-states of the operator $K_x$ with the eigen-value $k_x$. In view of the relation
$k_x = l_B^2 \,x_0$, we may as well say that  $|\,\Psi^{(C_1)}_{n, k_x} \rangle$ are
the eigen-states of the guiding-center operator $x_0$.) This observation
indicates purely theoretical nature of the operator $K_x$.
It would be certainly true that the pseudo momentum is a useful theoretical
concept, which enables us to understand the mutual relation between different gauge
choices. However, it is also true that there is
no direct relation between its gauge-invariant nature and observability.

A similar analyses can be carried out for the eigenstates
$| \,\Psi^{(C_2)}_{n, k_y} (x,y) \rangle$ and the corresponding momentum operators
$K^{can}_y$, $K^{mech}_y$, and $K_y$.
Since the general features are nothing different from the case of
$| \,\Psi^{(C_1)}_{n, k_x} (x,y) \rangle$,
except that the role of $x$ and $y$ coordinates is exchanged,
we do not repeat essentially the same analyses.

\noindent
\subsection{paths in circular coordinate}

\begin{figure}[ht]
\begin{center}
\includegraphics[width=6cm]{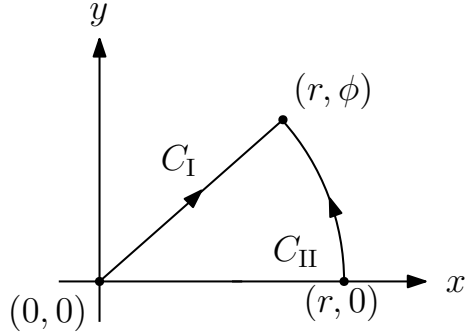}
\caption{The two paths $C_{\rm I}$ and $C_{\rm II}$ defined in
the circular coordinate system.}
\label{Fig:circular_path}
\end{center}
\end{figure}

Two simple choices of path  connecting the reference point 
$(r_0, \phi_0) \equiv (0,0)$ and the point $(r, \phi)$ in the circular coordinate
representation may be given by
\begin{eqnarray}
 C_{\rm I} \ &:& \ (0,0) \ \rightarrow \ (r, \phi) , \\
 C_{\rm II} \ &:& \ (0,0) \ \rightarrow \ (r, 0) \ \rightarrow \ (r, \phi) ,
\end{eqnarray}
as illustrated in Fig.\ref{Fig:circular_path}.

Selecting path $C_{\rm I}$, the relation between the original electron wave function
$\Psi (r, \phi)$ and the gauge-invariant wave function $\tilde{\Psi}^{(C_{\rm I})} (r, \phi)$
is given by
\begin{equation}
 \Psi (r, \phi) \ = \ U_{I} \,\tilde{\Psi}^{(C_{\rm I})} (r, \phi) , \label{Eq:Rep2}
\end{equation}
with
\begin{equation}
 U_{\rm I} \ \equiv \ e^{\,- \,i \,e \,\int_{C_{\rm I}} \,\bm{A} (\bm{x}) \cdot d \bm{x}}
 \ = \ e^{\,- \,i \,e \, \int_0^r \,A_r (r^\prime, \phi) \,d r^\prime} ,
\end{equation}
where $A_r (r,\phi)$ is the radial component of the vector potential.

On the other hand, with the choice of the path $C_{\rm II}$, we have
\begin{equation}
 \Psi (r, \phi) \ = \ U_{\rm II} \,\tilde{\Psi}^{(C_{\rm II})} (r, \phi) , \label{Eq:Rep1}
\end{equation}
with
\begin{equation}
 U_{\rm II} \ \equiv \ e^{\,- \,i \, e \,\int_{C_{\rm II}} \,\bm{A} (\bm{x}) \cdot d \bm{x}}
 \ = \ e^{\,- \,i \,e \,\left\{ \int_0^r \,A_r (r^\prime, 0) \,d r^\prime \ + \ 
 \int_0^\phi \,r \,A_\phi (r, \phi^\prime) \,d \phi^\prime \right\}},
\end{equation}
where $A_\phi (r, \phi)$ is the azimuthal component of the vector potential.

Again there is a nontrivial relation between the two operators $U_{\rm I}$ and
$U_{\rm II}$, which follows from the Stokes theorem. The relation is given by
\begin{equation}
 U_{\rm II}^{- \,1} \,U_{\rm I} \ = \ e^{\,i \,e \,
 \oint_{C_{\rm II} - C_{\rm I}} \,\bm{A} (\bm{x}) \cdot d \bm{x}} \ = \ 
 e^{\,i \,\frac{1}{2} \,e \,B \,r^2 \,\phi} . \label{Eq:U_I_U_II_relation}
\end{equation}
This identity gives a relation between the two forms of gauge-invariant electron
wave functions as
\begin{equation}
 \tilde{\Psi}^{(C_{\rm II})} (r, \phi) \ = \ e^{\,i \,\frac{1}{2} \,e \,B \,r^2 \,\phi} \,
 \tilde{\Psi}^{(C_{\rm I})} (r, \phi) . \label{Eq:WF_CII_CI}
\end{equation}
We shall come back to this relation when we discuss the gauge-invariant electron
wave functions corresponding to the path $C_{\rm II}$.

To get an insight into the implication of the two path choices, it is again
useful to examine the gauge-invariant photon fields corresponding
to the two paths $C_{\rm I}$ and $C_{\rm II}$. 
First, for the path choice $C_{\rm I}$, the gauge-invariant photon fields
defined by Eq.(\ref{Eq:GI_potential}) becomes
\begin{equation}
 \tilde{\bm{A}} (r, \phi) \ \equiv \ \tilde{\bm{A}}^{(C_{\rm I})} (r, \phi) \ = \ 
 \frac{1}{2} \,B \,r \,\bm{e}_\phi , \label{Eq:Atilde_C_I}
\end{equation}
which coincides with the gauge-potential in the symmetric gauge.
On the other hand, for the path choice $C_{\rm II}$, we find that
\begin{equation}
 \tilde{\bm{A}} (r, \phi) \ \equiv \ \tilde{\bm{A}}^{(C_{\rm II})} (r, \phi) \ = \ 
 - \,B \,r \,\phi \,\bm{e}_r , \label{Eq:Atilde_BB}
\end{equation}
which is the gauge potential in the so-called Bawin-Burnel gauge \cite{BB1983}. 
(We recall that the Bawin-Burnel gauge is a singular and {\it multi-valued} gauge.)
In any case, this simple analysis again indicates an intimate connection between
the choice of gauge and the choice of path. Even
stronger statement was made by Yang in a paper \cite{Yang1985}
titled ``Equivalence of path dependence and gauge dependence''.
According to his claim, a particular path choice is nothing but a particular
choice of gauge, i.e. they are totally equivalent. As we shall discuss below,
our conclusion is delicately different from his conclusion. 
Despite the intimate connection between the path choice and the gauge choice,
the choice of a particular path never prevents us from taking any other gauges.

Let us start our analysis with simpler path choice $C_{\rm I}$.
With this choice of path, the original Schr{\"o}dinger equation can be
transformed to the following equation for $\tilde{\Psi}^{(C_{\rm I})}$ :
\begin{equation}
 \tilde{H}_{\rm I} \, \tilde{\Psi}^{(C_{\rm I})} (r, \phi) \ = \ 
 E \, \tilde{\Psi}^{(C_{\rm I})} (r, \phi) , \label{Eq:H_I_tilde_eq}
\end{equation}
where 
\begin{equation}
 \tilde{H}_{\rm I} \ \equiv \ U_{\rm I}^{- \,1} \,H \,U_{\rm I} .
\end{equation}
After some lengthy calculation, we can show that
\begin{equation}
 \tilde{H}_{\rm I} \ = \ 
 \frac{\bm{p}^2}{2 \,m_e} \ + \ \frac{e}{m_e} \,\tilde{\bm{A}}^{(C_{\rm I})} \cdot \bm{p}
 \ - \ i \,\frac{e}{2 \,m_e} \,(\nabla \cdot \tilde{\bm{A}}^{(C_{\rm I})}) \ + \ 
 \frac{e^2}{2 \,m_e} \,\left(\tilde{\bm{A}}^{(C_{\rm I})} \right)^2 ,
\end{equation}
where $\tilde{\bm{A}}^{(C_{\rm I})}$ is given by (\ref{Eq:Atilde_C_I}).
This means that the transformed Hamiltonian $\tilde{H}_{\rm I}$ formally coincides
with the Landau Hamiltonian in the symmetric gauge. Explicitly, it takes
the form
\begin{equation}
 \tilde{H}_{\rm I} \ = \ 
 - \,\frac{1}{2 \,m_e} \,\frac{1}{r} \,\frac{\partial}{\partial r} \,
 \left( r \, \frac{\partial}{\partial r} \right) \ - \ 
 \frac{1}{2 \,m_e \,r^2} \,\frac{\partial^2}{\partial \phi^2} \ - \ 
 i \,\frac{e \,B}{2 \,m_e} \,\frac{\partial}{\partial \phi} \ + \ 
 \frac{e^2 \,B^2}{8 \,m_e} \,r^2 .
\end{equation}
Since this transformed Hamiltonian contains no $\phi$-dependent potential
term, the solution takes the following form :
\begin{equation}
 \tilde{\Psi}^{(C_{\rm I})} (r, \phi) \ = \ e^{\,i \,m \,\phi} \,R (r) , \label{Eq:C_I_separable}
\end{equation}
where $m$ is the eigenvalue of the canonical OAM operator $L^{can}_z$ as
\begin{equation}
 L^{can}_z \,\tilde{\Psi}^{(C_{\rm I})} (r, \phi) \ = \ m \,\tilde{\Psi}^{(C_{\rm I})} (r, \phi) .
 \label{Eq:eigen_eq_L_z}
\end{equation}
Putting (\ref{Eq:C_I_separable}) into (\ref{Eq:H_I_tilde_eq}), one finds that $R (r)$
satisfies the following equation : 
\begin{equation}
 \left\{ - \,\frac{1}{2 \,m_e} \,\frac{1}{r} \,\frac{d}{d r} \,
 \left( r \,\frac{d}{d r} \right) \ + \ \frac{1}{2 \,m_e \,r^2} \,
 \left( m + \frac{1}{2} \,e \,B \,r^2 \right)^2 \right\} \, R(r) \ = \ E \,R (r).
\end{equation}
The solution of this equation is well known. After all, we are led to the
following form of eigen-functions : 
\begin{equation}
 \tilde{\Psi}^{(C_{\rm I})}_{n,m} (r, \phi) \ = \ e^{\,i \,m \,\phi} \,R_{n,m} (r) ,
\end{equation}
with
\begin{equation}
 R_{n,m} (x,y) \ = \ \frac{1}{\sqrt{2 \,\pi}} \,\,
 N_{n,m} \,\left( \frac{r^2}{2 \,l_B^2} \right)^{|m|/2} \,
 e^{\,- \,\frac{r^2}{4 \,l_B^2}} \,
 L^{|m|}_{n - \frac{|m| + m}{2}} \left( \frac{r^2}{2 \,l_B^2} \right) . \label{Eq:R_nm}
\end{equation}
As a consequence, the original electron wave functions $\Psi (r, \phi)$ are given as
\begin{equation}
 \Psi (r, \phi) \ \equiv \ \Psi^{(C_{\rm I})}_{n,m} (r, \phi) \ = \ 
 U_{\rm I} \,\tilde{\Psi}^{(C_{\rm I})}_{n,m} (r, \phi) .
\end{equation}
Since the form of the original electron wave functions obtained in the above way 
depends on the chosen path $C_{\rm I}$ by construction, we have explicitly written
them as $\Psi^{(C_{\rm I})}_{n,m} (r, \phi)$.

As before, the eigen equation (\ref{Eq:eigen_eq_L_z}) for the canonical
OAM operator can be transformed back into the equation for the
original electron wave functions. It reads as
\begin{equation}
 U_{\rm I} \,L^{can}_z \,U_{\rm I}^{- \,1} \,\Psi^{(C_{\rm I})}_{n,m} (r, \phi) \ = \ m \,
 \Psi^{(C_{\rm I})}_{n,m} (r, \phi) .
\end{equation}
A simple manipulation shows that
\begin{equation}
 U_{\rm I} \,L^{can}_z \,U_{\rm I}^{- \,1} \ = \ - \,i \, \frac{\partial}{\partial \phi}
 \ + \ e \,r \,A_\phi \ - \ \frac{1}{2} \,e \,B \,r^2 .
\end{equation}
Note that the r.h.s. of this equation is nothing but the pseudo OAM operator
$L_z$ of Konstantinou and Moulopoulus \cite{Konstantinou2016},\cite{Konstantinou2017} : 
\begin{equation}
 L_z \ = \ (\bm{r} \times \bm{p})_z \ + \ e \,(\bm{r} \times \bm{A})_z \ - \ 
 \frac{1}{2} \,e \,B \,r^2 \ = \ 
 (\bm{r} \times \bm{\Pi})_z \ - \ \frac{1}{2} \,e \,B \,r^2 .
\end{equation}
To sum up, the electron wave functions obtained in the
above way are the simultaneous eigen-states of the operator
$L_z$ and the Landau Hamiltonian $H$ : 
\begin{eqnarray}
 H \,\Psi_{n,m}^{(C_{\rm I})} (r, \phi) &=& E_n \,\Psi_{n,m}^{(C_{\rm I})} (r, \phi) \ = \ 
 \left( n + \frac{1}{2} \right) \,\omega \,\Psi_{n,m}^{(C_{\rm I})} (r, \phi) , \\
 L_z \,\Psi_{n,m}^{(C_{\rm I})} (r, \phi) &=& m \,\Psi_{n,m}^{(C_{\rm I})} (r, \phi) .
\end{eqnarray}
If we use the relation (\ref{Eq:U_I_U_II_relation}), the above eigen-functions 
$\Psi^{(C_{\rm I})}_{n,m} (r, \phi)$ can also be expressed in the following form :  
\begin{eqnarray}
 \Psi^{(C_{\rm I})}_{n,m} (r, \phi) &=& U_{\rm_I} \,\tilde{\Psi}^{(C_{\rm I})}_{n,m} (r, \phi) 
 \ = \ U_{\rm II} \,e^{\,i \,\frac{1}{2} \,e \,B \,r^2 \,\phi} \,
 \tilde{\Psi}^{(C_{\rm I})}_{n,m} (r, \phi) \nonumber \\
 &=& e^{\,i \,m \,\phi} \,e^{\,i \,\frac{1}{2} \,e \,B \,r^2 \,\phi} \,
 e^{\,- \,i \,e \,\left\{ \int_0^r \,A_r (r^\prime, 0) \,d r^\prime \ + \ 
 \int_0^\phi \,r \,A_\phi (r, \phi^\prime) \,d \phi^\prime \right\}} \,R_{n,m} (r), 
 \label{Eq:LandauS_sol}
\end{eqnarray}
which reproduces the expression of Konstantinou and Moulopoulus obtained with
a different method.
We emphasize once again that these solutions are given independently of the choice of
the gauge potential $\bm{A}$.

To understand the physical implication of these eigen-states together with the
meaning of the gauge choice in the Landau problem, it is instructive to
investigate the function of three types of orbital angular momentum (OAM) operators,
i.e. the canonical OAM, the mechanical OAM, and the pseudo OAM on these
eigen-states.
Shown below are the effect of operation of the three
types of orbital
angular momentum (OAM) operators, $L^{can}_z$, $L^{mech}_z$, and $L_z$
on the states $| \Psi^{(C_{\rm I})}_{n,m} \rangle$, which are the simultaneous
eigen-states of the operator $L_z$ and the Landau Hamiltonian $H$ : 

\vspace{3mm}
\noindent
1) 1st Landau gauge \ : \ $\bm{A}_{L_1} = - \,B \,y \,\bm{e}_x$
\begin{eqnarray}
 L_z^{can} \,\Psi_{n,m}^{(C_{\rm I})} (r, \phi) &=&  
 \left(m + \frac{1}{2} \,e B r^2 \cos 2 \phi \right) \,\Psi_{n,m}^{(C_{\rm I})} (r, \phi) ,
 \label{Eq:1st_Landau_A} \\
 L_z^{mech} \,\Psi_{n,m}^{(C_{\rm I})} (r, \phi) &=& 
 \left(m + \frac{1}{2} \,e B r^2 \right) \,\Psi_{n,m}^{(C_{\rm I})} (r, \phi) , \\
 L_z \,\Psi_{n,m}^{(C_{\rm I})} (r, \phi) &=& m \,\Psi_{n,m}^{(C_{\rm I})} (r, \phi) .
\end{eqnarray}

\noindent
2) 2nd Landau gauge \ : \ $\bm{A}_{L_2} = B \,x \,\bm{e}_y$
\begin{eqnarray}
 L_z^{can} \,\Psi_{n,m}^{(C_{\rm I})} (r, \phi) &=&  
 \left(m - \frac{1}{2} \,e B r^2 \cos 2 \phi \right) \,\Psi_{n,m}^{(C_{\rm I})} (r, \phi) , 
 \label{Eq:2nd_Landau_A} \\
 L_z^{mech} \,\Psi_{n,m}^{(C_{\rm I})} (r, \phi) &=& 
 \left(m + \frac{1}{2} \,e B r^2 \right) \,\Psi_{n,m}^{(C_{\rm I})} (r, \phi) , \\
 L_z \,\Psi_{n,m}^{(C_{\rm I})} (r, \phi) &=& m \,\Psi_{n,m}^{(C_{\rm I})} (r, \phi) .
\end{eqnarray}

\noindent
3) symmetric gauge \ : \ $\bm{A}_{S} = \frac{1}{2} \,B \,r \,\bm{e}_\phi$
\begin{eqnarray}
 L_z^{can} \,\Psi_{n,m}^{(C_{\rm I})} (r, \phi) &=&  
 m \,\Psi_{n,m}^{(C_{\rm I})} (r, \phi) , \\
 L_z^{mech} \,\Psi_{n,m}^{(C_{\rm I})} (r, \phi) &=& 
 \left(m + \frac{1}{2} \,e B r^2 \right) \,\Psi_{n,m}^{(C_{\rm I})} (r, \phi) , \\
 L_z \,\Psi_{n,m}^{(C_{\rm I})} (r, \phi) &=& m \,\Psi_{n,m}^{(C_{\rm I})} (r, \phi) .
\end{eqnarray}

First, one notices that the result of operation of the
canonical OAM operator $L^{can}_z$ are all different, depending on the
three gauge choices. This is again a manifestation of the gauge-dependent
nature of the canonical OAM.
In contrast, reflecting the fact that the mechanical OAM operator
$L^{mech}_z$ and the pseudo OAM operator $L_z$ are both gauge-invariant, the
result of operation of these operators are all the same irrespectively of the gauge
choices. Note, however, that the operations of $L^{mech}_z$ and $L_z$ give
different answers.


\begin{table}[htb]
\caption{The expectation values of the three types of OAM operators,
$L^{can}_z, L^{mech}_z$, and $L_z$ 
on the states $| \Psi^{(C_{\rm I})}_{n,m} \rangle$.}
\label{Table:EV_Lz_eigenstate}
\vspace{4mm}
\renewcommand{\arraystretch}{1.2}
\begin{center}
\begin{tabular}{cccc}
 \hline
 \  & \  
 \ \ \ $\bm{A}_{L_1} = - B \,y \,\bm{e}_x$ \ \ \ &
 \ \ \ $\bm{A}_{L_2} = B \, x \,\bm{e}_x$ \ \ \ &
 \ \ \ $\bm{A}_S = \frac{1}{2} \,B \,r \,\bm{e}_\phi$ \ \ \ \\
 \hline
 \ $\langle L_z^{can} \rangle$ \ & 
 \ $m$ \ &
 \ $m$ \ & 
 \ $m$ \ \\
 \hline
 \ $\langle L_z^{mech} \rangle$ \ &  
 \ $2 \,n + 1$ \ &
 \ $2 \,n + 1$ \ & 
 \ $2 \,n + 1$ \ \\
 \hline
 \ $\langle L_z \rangle$ \ & 
 \ $m$ \ &
 \ $m$ \ & 
 \ $m$ \ \\
 \hline
\end{tabular}
\end{center}
\end{table}

To see it more transparently, let us compare in Table \ref{Table:EV_Lz_eigenstate}
the expectation values
of the three OAM operators $L^{can}_z$, $L^{mech}_z$, and $L_z$ between the
eigen-states $|\,\Psi^{(C_{\rm I})}_{n,m} \rangle$ with different gauge choices.
One observes that the expectation values of any of the three OAM
operators $L^{can}_z$, $L^{mech}_z$, and $L_z$ do not depend on
the choices of gauge. In particular, one sees that the expectation values
of the canonical OAM operator and the pseudo OAM operator are both $m$.
This coincidence is natural in the symmetric gauge. It is because
the pseudo OAM operator reduces to the canonical OAM operator with
this gauge choice. A little nontrivial is the observation that the expectation
value of the canonical OAM operator also coincides with $m$ in other two
gauges than the symmetric gauge. 
This should be contrasted with the fact that the expectation value of the
canonical momentum operators $K^{can}_x = p_x$ depends on the gauge choice.
This difference is related to the fact that there are two canonical momentum
operators $p_x$ and $p_y$, which cannot be diagonalized simultaneously, while
there is only one canonical OAM operator $L^{can}_z$.
As a matter of course, this fact alone cannot explain why the expectation value
of $L^{can}_z$ does not depend on the gauge choice.
In the case of the two Landau gauges, which we have investigated in the
present paper, the reason can be relatively easily understood.
It is because the matrix element of the terms containing 
$\cos 2 \,\phi$ in (\ref{Eq:1st_Landau_A}) and (\ref{Eq:2nd_Landau_A}) 
between the eigen-functions $|\,\Psi^{(C_I)}_{n,m} \rangle$ vanish after integration over
the azimuthal angle $\phi$. A natural question is what happens with other
gauges than the two Landau gauges. We found that the expectation value
of the canonical OAM operator between the  eigenstates  $|\,\Psi^{(C_I)}_{n,m} \rangle$
is given by $m$ in arbitrary single-valued (or regular) gauges. An essential
ingredient leading to this result is the periodicity of the wave functions
$\Psi^{(C_I)}_{n,m} (r, \phi)$ as well as the gauge field configuration
$\bm{A} (r, \phi)$ with respect to the angle $\phi$ with the $2 \,\pi$ period.
To our best knowledge, this fact has not been shown explicitly before.
We therefore think it useful to give its proof in Appendix D.

Turning to the expectation value of the mechanical OAM operator,
we see that it is given by $2 \,n + 1$, with $n$ being the familiar Landau quantum
number, irrespectively of the gauge choice.
Since this quantum number appears in the expression of the Landau energy
levels $E_n = \left( n + \frac{1}{2} \right) \,\omega$, it clearly
corresponds to an observable.
This is not the case for the quantum number $m$.
As is well-known, each Landau level with a fixed quantum number $n$ has
infinitely many degeneracies with different values of $m$.
This means that, at least in the Landau problem, the quantum number $m$
and therefore the canonical OAM as well as the pseudo OAM are not such
quantities, which have direct connection with observables.

Also very important to recognize is the fact that the simultaneous eigen-states of $L_z$ and $H$,
which we have discussed above, do not reduce to the standardly-known
eigen-states in the Landau gauge even in the limit of Landau gauge fixing
in the above gauge-invariant formulation.
To confirm it, we recall that, in the gauge-invariant formulation, 
the simultaneous eigen-states of
$L_z$ and $H$ are given as
\begin{equation}
 \Psi_{n,m}^{(C_{\rm I})} (r, \phi) \ = \ U_{\rm I} \,e^{\,i \,m \,\phi} 
 \,R_{n,m} (r) \ = \ 
 e^{\,i \,m \,\phi} \,e^{\,- \,i \,e \,
 \int_0^r \,A_r (r^\prime, \phi) \,d r^\prime} \,R_{n,m} (r) ,
\end{equation}
where $R_{m,n} (r)$ is given by (\ref{Eq:R_nm}).
With the choice of the symmetric gauge, which amounts to taking
$A_r \, = \, 0, A_\phi \ = \ \frac{1}{2} \,B \,r$,
the above eigen-functions reduce to
\begin{equation}
 \Psi_{n,m}^{(C_{\rm I})} (r, \phi) \ \rightarrow \ e^{\,i \,m \,\phi} \,R_{n,m} (r) ,
\end{equation}
which is nothing but the standard eigen-functions of the Landau Hamiltonian
in the symmetric gauge.
On the other hand, in the limit of taking the 2nd type of Landau gauge,
which amounts to taking
$A_r \, = \, r \,B \,\cos \phi \,\sin \phi, A_\phi \, = \, r \,B \,\cos^2 \phi$,
the eigen-functions reduces to
\begin{equation}
 \Psi_{n,m}^{(C_{\rm I})} (r, \phi) \ \rightarrow \ e^{\,i \,m \,\phi} \,
 e^{\,- \,i \,\frac{1}{2} \,e \,B \,x \,y} \,R_{n,m} (r) .
\end{equation}
Clearly, they do not coincides with the standard eigen-functions
of the Landau Hamiltonian in the Landau gauge. The truth is that,
as we have already discussed in sect \ref{Section:s1}, the r.h.s. of
the above equation is proportional to a particular
superposition of the standard eigen-functions in the Landau gauge.
Namely, the eigen-function $\Psi_{n,m}^{(C_{\rm I})} (r, \phi)$ in the gauge-invariant
formulation reduces in the Landau gauge limit to the following functions : 
\begin{equation}
 \Psi_{n,m}^{(C_{\rm I})} (r, \phi) \ \rightarrow \ e^{\,- \, i \,\frac{1}{2} \,e \,B \,x \,y} \,
 \int \,d k_y \,\,U_{n,m} (x_0) \,\Psi_{n,k_y} (x,y) ,
\end{equation}
where $\Psi_{n,k_y} (x,y)$ is the standard eigen-functions of the Landau
Hamiltonian in the 1st Landau gauge, while the weight function
$U_{n,m} (x_0)$ of the superposition is given as the matrix element
of the gauge transformation
operator between the standard eigen-states in the Landau gauge and the
standard eigen-functions in the symmetric gauge as shown in sect \ref{Section:s1}.

\subsection{A short summary on the issue of gauge choice in the Landau problem}

The delicacy of the gauge choice in the Landau problem should be clear by now.
In the standard treatment, the choice of gauge and that of coordinate system
is inseparably connected. This is so because, if we choose the Landau gauge, the
rectangular-coordinate description is more natural and economical, since
the gauge potential in this Hamiltonian depends on either of coordinate $x$ or $y$
only.
On the other hand, if the symmetric gauge is chosen, the gauge potential
has an axial symmetry and it has azimuthal component only. Therefore,
the circular coordinate choice is more convenient to work with.

For getting still deeper insight into the delicacy of the gauge choice 
in the Landau problem, more powerful is the gauge-invariant formulation
of the same problem.
For the Landau problem, two different types of gauge-invariant
formulation are possible.
The first is the gauge-invariant formulation of Konstantinou
and Moulopoulos \cite{Konstantinou2016},\cite{Konstantinou2017} or of 
Haugset et al. \cite{Haugset1993}. In this formulation, what plays an
important role is the existence of the two pseudo-momentum
operator $\bm{K}$ and the pseudo OAM operator $L_z$, which
are both gauge-invariant objects. A remarkable fact is that
we can obtain the simultaneous eigen-states of one of these operators and
the Landau Hamiltonian without fixing gauge.
This is assured by the fact that the gauge-invariant operator $\bm{K}$ as well as
the gauge-invariant operator $L_z$ commute with the Landau
Hamiltonian \cite{Konstantinou2016},\cite{Konstantinou2017},
\begin{equation}
 [\bm{K}, H] \ = \ 0, \ \ \ [L_z, H] \ = \ 0.
\end{equation}
These operators are not mutually commutable, however,
\begin{equation}
 [K_x, K_y] \ \neq \ 0, \ \ \ [K_x, L_z] \ \neq \ 0, \ \ \ 
 [K_y, L_z] \ \neq \ 0.
\end{equation}
The possibilities are therefore to look for simultaneous eigen-states of $K_x$
and $H$, or $K_y$ and $H$, or $L_z$ and $H$.
The first two choices give eigen-functions, which can be expressed in terms of
Hermite polynomials, while the third choice gives eigen-functions, which can be
expressed with associated Laguerre polynomials.
As emphasized by Haugset et al.~\cite{Haugset1993}, 
``Which of eigen-functions one
wants to work with is basically not equivalent to the choice of gauge, but rather
which operator one wants to have diagonalized.''
An advantage of working in the third type of eigen-function is that
the axial symmetry of the problem is maintained and the meaning of the
angular momentum operators becomes clear.
In fact, as we have seen, to give a clear meaning to these quantities,
we had to consider a superposition of the eigen-states if we stick
to working in the first two types of formulations.

Actually, we have solved the Landau problem based on another
gauge-invariant formulation, i.e. DeWitt's formalism.
This formulation has broader utility than the above-mentioned one,
since the existence of the {\it local} and gauge-invariant
pseudo momenta and pseudo OAM is not assumed as a prerequisite of
formulation. (As we shall see in later section, this feature is important
for our discussion of the nucleon spin decomposition problem,
because no analogous local quantities like the ones $\bm{K}$ and $L_z$
are not known to exist in this case.)
Instead, what is essential in DeWitt's formalism is the freedom
in the choice of path. As we have shown, with the choice of the path
$C_1$ in the rectangular coordinate, for example, the gauge-invariant
photon field reduces to the vector potential in the 1st Landau gauge. 
In this sense, the choice of the path $C_1$ is
intimately connected with the choice of the 1st Landau gauge.
Nonetheless, this choice of path is not totally equivalent to taking
the 1st Landau gauge. Within the gauge-invariant formulation of
DeWitt, we still have a freedom to take other gauges like
the 2nd Landau gauge and the symmetric gauge.
A natural question is therefore what is special with the
choice of the path $C_1$. The answer is already clear. It is the realization
of translational symmetry with respect to the $x$-axis.
In fact, as one can see clearly from the fourth column of
Table \ref{Table:EV_Kx_eigenstate}, this symmetry manifests itself in the
gauge-choice independence of the expectation value of the pseudo-momentum
operator $K_x$ or the g.i.c. momentum operator in our
terminology.

Similarly, if we take the path $C_I$ in the cylindrical
coordinate, it amounts to respecting the axial symmetry
or the rotational symmetry around the $z$-axis.
Again, it is not equivalent to taking the symmetric gauge,
since we can still choose other gauges like the two
Landau gauges. In this case, the axial symmetry manifests
in the gauge-choice independence of the expectation value
of the pseudo OAM operator or the g.i.c. OAM in our
terminology, as can be seen from the fourth column of
Table \ref{Table:EV_Lz_eigenstate}.
In this way, we arrive at an important conclusion.
Choosing a particular path is equivalent to deciding what symmetry
is to be respected in the physics of our interest.
We emphasize that we have proven this fact based on a concrete example,
which makes full use of the analytically-obtained wave
functions of the Landau problem within DeWitt's formalism.

\subsection{The choice of path $C_{II}$ and the multi-valued gauge}

Before ending this section, for the sake of completeness, we briefly
discuss what happens if we choose another path $C_{\rm II}$ in the
cylindrical coordinate for defining the gauge-invariant electron wave
functions. As already pointed out, the gauge-invariant wave functions
$\tilde{\Psi}^{(C_{\rm II})} (r, \phi)$ corresponding to the path $C_{\rm II}$ are related to
the wave functions $\tilde{\Psi}^{(C_{\rm I})} (r, \phi)$ by Eq. (\ref{Eq:WF_CII_CI}),
i.e. as $\tilde{\Psi}^{(C_{\rm II})} (r, \phi) \, = \, e^{\,\frac{1}{2} \,i \,e \,B \,r^2 \,\phi} \,
\tilde{\Psi}^{(C_{\rm I})} (r, \phi)$.
Since $\tilde{\Psi}^{(C_{\rm I})} (r, \phi)$ is a periodic function of $\phi$ with 
a period $2 \,\pi$, i.e.
\begin{equation}
 \tilde{\Psi}^{(C_{\rm I})} (r, \phi + 2 \,\pi) \ = \ \tilde{\Psi}^{(C_{\rm I})} (r, \phi) ,
\end{equation}
it follows that the wave functions $\tilde{\Psi}^{(C_{\rm II})} (r, \phi)$ satisfy the
following unusual boundary condition,
\begin{equation}
 \tilde{\Psi}^{(C_{\rm II})} (r, \phi + 2 \,\pi) \ = \ 
 e^{\,i \,\pi \,e \,B \,r^2} \,
 \tilde{\Psi}^{(C_{\rm II})} (r, \phi) ,
\end{equation}
which means that they are multi-valued functions with respect to the
variable $\phi$. We point out that
this is consistent with the fact that the gauge-invariant photon field
corresponding to the path $C_{\rm II}$ is given by (\ref{Eq:Atilde_BB}), which is just the
vector potential in the (multi-valued) Bawin-Burnel (BB) gauge.

In any case, using the relation (\ref{Eq:WF_CII_CI}), the electron wave-functions
corresponding to the path choice $C_{\rm II}$ can be readily obtained as
\begin{eqnarray}
 \Psi^{(C_{\rm II})}_{n,m} (r, \phi) &=& U_{\rm II} \,\,
 \tilde{\Psi}^{(C_{\rm II})}_{n,m} (r \,\phi)
 \ = \ U_{\rm II} \,\,e^{\,i \,\frac{1}{2} \,e \,B \,r^2 \,\phi} \,
 \tilde{\Psi}^{(C_{\rm I})}_{n,m} (r \,\phi) \nonumber \\
 &=& U_{\rm II} \,e^{\,i \,\frac{1}{2} \,e \,B \,r^2 \,\phi} \,
 e^{\,i \,m \,\phi} \,R_{n,m} (r) \nonumber \\
 &=& e^{\,i \,\left( m \, + \, \frac{1}{2} \,e \,B \,r^2 \right) \,\phi} \,
 e^{\,- \,i \,e \,\left\{ \int_0^r \,A_r (r^\prime, 0) \,d r^\prime \ + \ 
 \int_0^\phi \,r \,A_\phi (r, \phi^\prime) \,d \phi^\prime \right\}} \,R_{n,m} (r) ,
\end{eqnarray}
where $R_{n,m} (r)$ is given by (\ref{Eq:R_nm}).
Again, one may be interested in the result of operation of the three types
of orbital angular momentum operators, i.e. $L^{can}_z$, $L^{mech}_z$,
and $L_z$ on the states $|\,\Psi^{(C_{\rm II})}_{n,m} \rangle$. 
Using the circular coordinate representation of these OAMs given by
\begin{eqnarray}
 L^{can}_z &=& - \,i \,\frac{\partial}{\partial \phi}, \\
 L^{mech}_z &=& - \,i \,\frac{\partial}{\partial \phi} \ + \ e \,r \,A_\phi , \\
 L_z &=& - \,i \,\frac{\partial}{\partial \phi} \ + \ e \,r \,A_\phi \ - \ 
 \frac{1}{2} \,e \,B \,r^2 , 
\end{eqnarray}
one is led to the following answer :

\vspace{3mm}
\noindent
1) 1st Landau gauge \ : \ $\bm{A}_{L_1} = - \,B \,y \,\bm{e}_x$
\begin{eqnarray}
 L_z^{can} \,\Psi_{n,m}^{(C_{\rm II})} (r, \phi) &=&  
 \left(m + \frac{1}{2} \,e \,B \,r^2 \,\cos 2 \phi \right) \,\Psi_{n,m}^{(C_{\rm II})} (r, \phi), 
 \\
 L_z^{mech} \,\Psi_{n,m}^{(C_{\rm II})} (r, \phi) &=& 
 \left(m + \frac{1}{2} \,e \,B \,r^2 \right) \,\Psi_{n,m}^{(C_{\rm II})} (r, \phi), \\
 L_z \,\Psi_{n,m}^{(C_{\rm II})} (r, \phi) &=& m \,\Psi_{n,m}^{(C_{\rm II})} (r, \phi) .
\end{eqnarray}

\noindent
2) 2nd Landau gauge \ : \ $\bm{A}_{L_2} = B \,x \,\bm{e}_y$
\begin{eqnarray}
 L_z^{can} \,\Psi_{n,m}^{(C_{\rm II})} (r, \phi) &=&  
 \left(m - \frac{1}{2} \,e \,B \,r^2 \,\cos 2 \phi \right) \,\Psi_{n,m}^{(C_{\rm II})} (r, \phi), 
 \\
 L_z^{mech} \,\Psi_{n,m}^{(C_{\rm II})} (r, \phi) &=& 
 \left(m + \frac{1}{2} \,e \,B \,r^2 \right) \,\Psi_{n,m}^{(C_{\rm II})} (r, \phi), \\
 L_z \,\Psi_{n,m}^{(C_{\rm II})} (r, \phi) &=& m \,\Psi_{n,m}^{(C_{\rm II})} (r, \phi) .
\end{eqnarray}

\noindent
3) symmetric gauge \ : \ $\bm{A}_{S} = \frac{1}{2} \,B \,r \,\bm{e}_\phi$
\begin{eqnarray}
 L_z^{can} \,\Psi_{n,m}^{(C_{\rm II})} (r, \phi) &=&  
 m \,\Psi_{n,m}^{(C_{\rm II})} (r, \phi), \\
 L_z^{mech} \,\Psi_{n,m}^{(C_{\rm II})} (r, \phi) &=& 
 \left(m + \frac{1}{2} \,e \,B \,r^2 \right) \,\Psi_{n,m}^{(C_{\rm II})} (r, \phi), \\
 L_z \,\Psi_{n,m}^{(C_{\rm II})} (r, \phi) &=& m \,\Psi_{n,m}^{(C_{\rm II})} (r, \phi) .
\end{eqnarray}

\noindent
4) Bawin-Burnel (BB) gauge \ : \ $\bm{A}_{BB} = - \,B \,r \,\bm{e}_r$
\begin{eqnarray}
 L_z^{can} \,\Psi_{n,m}^{(C_{\rm II})} (r, \phi) &=&  
 \left(m + \frac{1}{2} \,e \,B \,r^2 \right) \,\Psi_{n,m}^{(C_{\rm II})} (r, \phi), \\
 L_z^{mech} \,\Psi_{n,m}^{(C_{\rm II})} (r, \phi) &=& 
 \left(m + \frac{1}{2} \,e \,B \,r^2 \right) \,\Psi_{n,m}^{(C_{\rm II})} (r, \phi), \\
 L_z \,\Psi_{n,m}^{(C_{\rm II})} (r, \phi) &=& m \,\Psi_{n,m}^{(C_{\rm II})} (r, \phi) .
\end{eqnarray}

One confirms that the results of operation of $L^{mech}_z$ and $L_z$
do not depend on the gauge-field configuration, i.e. they are the same
for all of the choices $\bm{A}_{L_1}$, $\bm{A}_{L_2}$, $\bm{A}_S$, and $\bm{A}_{BB}$.
However, this is not the case for the canonical OAM operator $L^{can}_z$, which
is a gauge-variant operator. 


\vspace{3mm}
\begin{table}[h]
\caption{The expectation values of the three types of OAM operators,
$L^{can}_z, L^{mech}_z$, and $L_z$ 
on the states $| \Psi^{(C_{\rm II})}_{n,m} \rangle$.}
\label{Table:EV_Lz_eigenstate_BB}
\vspace{4mm}
\renewcommand{\arraystretch}{1.2}
\begin{center}
\begin{tabular}{ccccc}
 \hline
 \  & \  
 $\bm{A}_{L_1} = - B \,y \,\bm{e}_x$ &
 $\bm{A}_{L_2} = B \, x \,\bm{e}_x$ &
 $\bm{A}_S = \frac{1}{2} \,B \,r \,\bm{e}_\phi$ &
 $\bm{A}_{BB} \ = \,- \,B \,r \,\phi \,\bm{e}_r$ \\
 \hline
 \ $\langle L_z^{can} \rangle$ \ & 
 \ $m$ \ &
 \ $m$ \ & 
 \ $m$ \ &
 \ $2 \,n + 1$ \\
 \hline
 \ $\langle L_z^{mech} \rangle$ \ &  
 \ $2 \,n + 1$ \ &
 \ $2 \,n + 1$ \ &
 \ $2 \,n + 1$ \ &
 \ $2 \,n + 1$ \ \\
 \hline
 \ $\langle L_z \rangle$ \ & 
 \ $m$ \ &
 \ $m$ \ &
 \ $m$ \ & 
 \ $m$ \ \\
 \hline
\end{tabular}
\end{center}
\end{table}

Also interesting to see is the expectation values of the three OAM
operators between the eigenstates $|\,\Psi^{(C_{\rm II})}_{n,m} \rangle$.
They are shown in Table \ref{Table:EV_Lz_eigenstate_BB}.
One finds that the expectation values of $L^{mech}_z$ are all $2 \,n + 1$
irrespectively of the choices of the gauge potential.
On the other hand, the expectation values of $L_z$ are all $m$,
independently of the choices of gauge potential.
However, the expectation values of $L^{can}_z$ is $2 \,n + 1$ in the BB gauge,
whereas they are $m$ in other three gauges.
Thus, for the canonical OAM operator, we find that even the
expectation values are generally dependent on the gauge choice.
Also very interesting is the fact that the expectation values of
$L^{can}_z$ and that of $L_z$ do not coincide in the BB gauge.
The gauge-choice independence of the pseudo-OAM operator
$L_z$ can be interpreted as showing its unique theoretical nature
as the generator of rotation around the 
$z$-axis \cite{Konstantinou2016},\cite{Konstantinou2017}.

\vspace{4mm}
\section{On the concept of gauge-invariant canonical orbital angular momentum}
\label{Section:s4}

We have reconfirmed that the pseudo momentum $\bm{K}$ and
the pseudo OAM $L_z$ defined by
\begin{eqnarray}
 \bm{K} &\equiv& \ \bm{\Pi} \ + \ e \,\bm{r} \times \bm{B}, \\
 L_z &\equiv& (\bm{r} \times \bm{\Pi})_z \ - \ \frac{1}{2} \,e \,B \,r^2,
\end{eqnarray}
play important roles in the gauge-invariant formulation of the 
Landau problem \cite{Konstantinou2016},\cite{Konstantinou2017}.
A remarkable features of these quantities are that
they are apparently gauge-invariant and besides they reduce to the
canonical momentum and canonical OAM in suitable gauges. 
This reminds us of the concept of gauge-invariant canonical (g.i.c.) momentum
and the gauge-invariant canonical (g.i.c.) OAM, which were
advocated by Chen et al., and intensively discussed in the context
of the nucleon spin decomposition problem as well as the
photon spin decomposition problem \cite{Chen2008},\cite{Chen2009}. 
Their basic idea starts with decomposing the gauge field into 
what-they-call the ``physical'' component and the ``pure-gauge'' component as
\begin{equation}
 \bm{A} \ = \ \bm{A}_{phys} \ + \ \bm{A}_{pure}.
\end{equation}
If one restricts to the case of quantum electrodynamics, which is an abelian
gauge theory, their proposal reduces to the decomposition of the total photon field
(or the vector potential) into the following two components,
\begin{equation}
 \bm{A} \ = \ \bm{A}_\perp \ + \ \bm{A}_\parallel ,
\end{equation}
where $\bm{A}_\perp$ and $\bm{A}_\parallel$ are respectively the familiar transverse
and longitudinal components of the photon field, satisfying the conditions
$\nabla \cdot \bm{A}_\perp = 0$
and $\nabla \times \bm{A}_\parallel = 0$. Once the Lorentz frame is fixed,
this transverse-longitudinal decomposition is known to be unique
owing to the Helmholtz theorem, provided that the
vector potential $\bm{A}$ damps fast enough at the spatial infinity.
Also noteworthy is the gauge transformation property of the two components
$\bm{A}_\perp$ and $\bm{A}_\parallel$. Under general gauge transformations
$\bm{A} \rightarrow \bm{A} + \nabla \chi$, it can be shown that they transform
as $\bm{A}_\perp \rightarrow \bm{A}_\perp$ and $\bm{A}_\parallel \rightarrow
\bm{A}_\parallel + \nabla \chi$. That is, the transverse component is
gauge-invariant, while the longitudinal component carries the residual
gauge degrees of freedom. From this fact, it immediately follows
that the operators defined by
\begin{eqnarray}
 \bm{p}^{g.i.c.} &\equiv& \bm{p} \ + \ e \, \bm{A}_\parallel ,\\
 L^{g.i.c.}_z &\equiv& [\bm{r} \times (\bm{p} + e \,\bm{A}_\parallel)]_z,
\end{eqnarray}
transform just in the same way as
\begin{eqnarray}
 \bm{p}^{mech} &=& \bm{p} \ + \ e \,\bm{A} , \\
 L^{mech}_z &=& [\bm{r} \times (\bm{p} \ + \ e \,\bm{A})]_z ,
\end{eqnarray}
which in turn means that $\bm{p}^{g.i.c.}$ as well as $L^{g.i.c.}_z$ are both
gauge-invariant. Besides, if the gauge degrees of freedom are eliminated,
which allows to set $\bm{A}_\parallel = 0$, they clearly reduce to the ordinary
canonical momentum and canonical OAM operators. 
For this reason, these quantities are often
called the gauge-invariant canonical (g.i.c.) momentum and the
gauge-invariant canonical (g.i.c.) OAM in the 
literatures \cite{Review_LL14},\cite{Review_LL16}.

As can be easily verified, the gauge potential $\bm{A}_S$ in the symmetric gauge
satisfies the transverse condition $\nabla \cdot \bm{A}_S = 0$.
(Since this condition is just the same as the Coulomb gauge condition,
the symmetric gauge is sometimes called the Coulomb gauge
in some paper \cite{Greenschields2014}.) 
Then, one might be tempted to identity the gauge potential in
the symmetric gauge with the transverse component of the vector potential.
With this identification, we have
\begin{equation}
 (\bm{r} \times \bm{A}_\perp)_z \ = \ (\bm{r} \times \bm{A}_S)_z
 \ = \ \frac{1}{2} \,B \,r^2 .
\end{equation}
As a consequence, the pseudo OAM operator $L_z$ reduces to
\begin{eqnarray}
 L_z &=& (\bm{r} \times \bm{p})_z \ + \ e \,(\bm{r} \times \bm{A}_\perp)_z
 \ + \ e \,(\bm{r} \times \bm{A}_\parallel)_z \ - \ 
 \frac{1}{2} \,e \,B \,r^2 \nonumber \\
 &=& \left[\bm{r} \times \left( \bm{p} \ + \ e \,\bm{A}_\parallel \right) \right]_z
 \ = \ \left( \bm{r} \times \bm{D}_{pure} \right)_z  \ = \ L^{g.i.c}_z ,
\end{eqnarray}
with $\bm{D}_{pure} \equiv \bm{p} \,+ \,e \,\bm{A}_\parallel$ being the so-called
pure-gauge covariant derivative.
This last quantity is nothing but the gauge-invariant canonical OAM
of Chen et al.

Unfortunately, the condition of the Helmholtz theorem is not
satisfied in the Landau problem, since the magnetic field spreads
over the whole plane.
As a consequence, the transverse-longitudinal decomposition cannot
be carried out uniquely and the above identification is also not
unique. In fact, one can easily verify that
the vector potentials $\bm{A}_{L_1}$, $\bm{A}_{L_2}$, and $\bm{A}_S$
in the 1st and 2nd Landau gauges and
the symmetric gauge, all satisfy the transverse condition, i.e.
$\nabla \cdot \bm{A}_{L_1} = \nabla \cdot \bm{A}_{L_2} = 
\nabla \cdot \bm{A}_S = 0$.
Still, as we have discussed in sect \ref{Section:s1}, 
if one wants to analyze the
orbital angular momentum of the electron, the symmetric gauge
appears to be the most natural and convenient choice. 
(As a matter of course, we expect that 
the genuine physical observables like the Landau levels are
independent of the choice of gauges.)
This is obviously related to the fact that, even though the
uniform magnetic field in the Landau problem does not have
any preferred symmetry axis, the motion
of the electron in this magnetic field is a circular motion around
some center, which means that there is an apparent cylindrical symmetry in physics.

What if there is an axial symmetry in the magnetic
field distribution from the outset, then ?
In such a situation, the choice of symmetric gauge seems
almost mandatory, or at least most convenient. 
As an example, let us consider an
infinitely long solenoid along the $z$-direction with $R$ being the radius
of the cross section of it.
In this setting, there is a uniform magnetic field only inside
the solenoid, while there is no magnetic field outside it, such that
\begin{equation}
 \bm{B} (\bm{x}) \ = \ B_0 \,\theta (R - r) \,\bm{e}_z .
\end{equation}
Li et al. argued that the transverse component $\bm{A}_\perp$ of
the vector potential for this magnetic field can uniquely be determined
by the Helmholtz theorem \cite{Li-Wang2012}. 
According to the Helmholtz theorem, 
$\bm{A}_\perp$ can formally be expressed as
\begin{eqnarray}
 \bm{A}_\perp (\bm{x}) \!\!\! &=& \!\!\!
 \nabla \times \frac{1}{4 \,\pi} \,\int \,d^3 x^\prime \,
 \frac{\nabla^\prime \times \bm{A} (\bm{x}^\prime)}
 {|\bm{x} - \bm{x}^\prime|} \nonumber \\
 \!\!\! &=& \!\!\!
 \nabla \times \frac{1}{4 \,\pi} \,\int \,d^3 x^\prime \,
 \frac{\bm{B} (\bm{x}^\prime)}
 {|\bm{x} - \bm{x}^\prime|}
 \ = \ 
 \nabla \times \frac{B_0}{4 \,\pi} \,\,\bm{e}_z \,
 \int_{|\bm{x}^\prime| \leq R} \,
 \frac{d^3 x^\prime}{|\bm{x} - \bm{x}^\prime|} . \hspace{6mm}
 \label{Eq:A_perp}
\end{eqnarray}
Actually, for an infinitely long solenoid, the above integral is
logarithmically divergent.
To regulate the integral, Li et al. first consider 
a solenoid with large but finite length $L$ \cite{Li-Wang2012}.
For this solenoid, they showed that
\begin{equation}
 \int_{|\bm{x}^\prime| < R} \,\frac{d^3 x^\prime}
 {|\bm{x} - \bm{x}^\prime|} \ = \ \pi \,R^2 \,\ln 4 \,L^2 \ + \ 
 \pi \,R^2 \,\ln R^2 \ - \ \pi R^2 \ + \ \pi \,r^2 \ + \ 
 O \left( \frac{1}{L^2} \right) ,
\end{equation}
for $r < R$, and
\begin{equation}
 \int_{|\bm{x}^\prime| < R} \,\frac{d^3 x^\prime}
 {|\bm{x} - \bm{x}^\prime|} \ = \ \pi \,R^2 \,\ln 4 \,L^2 \ + \ 
 \pi \,R^2 \,\ln r^2 \ + \ 
 O \left( \frac{1}{L^2} \right) ,
\end{equation}
for $r > R$. Inserting these expressions into (\ref{Eq:A_perp}), and
then taking the $L \rightarrow \infty$ limit, they obtain
\begin{equation}
 \bm{A}_\perp (\bm{x}) \ = \ 
 \left\{ \begin{array}{ll}
 \ \frac{1}{2} \,B \,r \,\bm{e}_\phi \ & \ (r \leq R) , \\
 \ \frac{1}{2} \,\frac{B \,R^2}{r} \,\bm{e}_\phi \ & \ (r > R) , \\
 \end{array} \right. 
\end{equation}
which confirms that the Helmholtz theorem uniquely
fix the transverse component of the vector potential.
Very interestingly, if one further takes the limit $R \rightarrow \infty$, which
amounts to considering a uniform magnetic field spread over the whole
$x$-$y$ plane, one obtains
\begin{equation}
 \bm{A}_\perp (\bm{x}) \ \rightarrow \ \frac{1}{2} \,B \,r \,\bm{e}_\phi .
\end{equation}
This is nothing but the gauge potential of the symmetric gauge in the
Landau problem.

Our analysis shows that the canonical OAM is an important
quantity that characterizes the quantum mechanical
eigen-states of a Hamiltonian with cylindrical symmetry.
Nevertheless, it does not correspond to any observable
at least in the Landau problem.
However, we refrain from concluding that the canonical OAM
can never be observed. One of the reasons is that we are aware
of the recent research progress, which suggests possible
observability of the canonical OAM \cite{Schattschneider2014},\cite{Bliokh2012}. 
The point is to consider the motion of electron along the
direction of the uniform magnetic field, which was neglected in
most analyses of the Landau problem in the past.
From the technical viewpoint, the idea is motivated by the development
of the new experimental technique, i.e. the generation of the
so-called {\it electron vortex beam} by utilizing the transmission
electron microscopes (TEM). 
They proposed to use this electron vortex beam running along
the direction of the magnetic field, which is supposed to be
practically uniform as compared with the size of the electron
vortex beam. According to them, this setting enables real-space
elucidation of individual Landau level, which is characterized  by two
quantum numbers, i.e. the Landau quantum number $n$ and
the eigen-value $m$ of the canonical OAM operator.
In fact, they showed that, while propagating along the direction of
the magnetic field, the Landau electrons receive characteristic
rotation with three different angular velocities, depending on
the eigen-value $m$ of the canonical OAM.
This rotation during the propagation along the
$z$-direction can be measured experimentally by using
statistical averaging over many identical single-particle events.
This then enables to get direct relation between the
canonical OAM and an observable.
It is a remarkable findings, in the sense that it sheds light on
the so-far hidden $m$-dependent rotational dynamics of the quantum
Landau states for the first time. However, it also appears to
bring about a new question on the interpretation of the
widely-accepted fundamental
principle of physics, i.e. the {\it gauge principle}.
In fact, according to the gauge principle,
observables must be gauge-invariant, but the canonical OAM is
not a gauge-invariant quantity.

\section{Remarks on the issue of gauge choice in the nucleon
spin decomposition problem}
\label{Section:s5}

In the discussion of the gauge-invariant formulation of the Landau problem,
we have pointed out that there are in principle infinitely many choices of path
for defining the gauge-invariant electron and photon fields and besides that
there is an intimate connection between the choice of {\it path}
and the choice of {\it gauge}.
Once, Belinfante showed that, by averaging over path-dependent potential over
the direction of all straight line paths at constant time converging to the point
where the gauge potential is to be calculated, one is led to the gauge potential
in the Coulomb gauge \cite{Belinfante1962}. 
On the other hand, Rohrlich and Strocchi applied a
similar averaging procedure over a covariant path, thereby obtaining the
gauge potential in the Lorentz gauge \cite{Rhorlich-Strocchi1965}. 
Since the Coulomb gauge is one of
the most important gauges, which falls into the class of the so-called
physical gauges, we briefly introduce Belinfante's argument \cite{Belinfante1962}.
(The physical gauge is a particular class of gauges, which maintains only two physical
degrees of freedom of the massless vector field. 
The greatest advantage of the physical gauge is that they
do not need to introduce the Hilbert space with indefinite metric, which is
unavoidable in the quantization with the covariant gauge choice
like the Feynman gauge or the Landau gauge.) 

Belinfante's argument starts with the general expression of the gauge-invariant
photon field given as
\begin{equation}
 \tilde{A}_\mu (x) \ = \ \int_{- \,\infty}^0 \,F_{\nu \sigma} (z) \,
 \frac{\partial z^\nu}{\partial x^\mu} \,\frac{\partial z^\sigma}{\partial \xi} \,
 d \xi , \label{Eq:Atilde_Bel}
\end{equation}
where $z^\nu (x,\xi)$ represents a point on the line toward
$x$, with $\xi$ being a
parameter chosen to satisfy the following boundary condition :
\begin{equation}
 z^\nu (x,0) \ = \ x^\nu \ \ \ \mbox{and} \ \ \ 
 z^\nu (x, - \,\infty) \ = \ \mbox{spatial infinity} .
\end{equation}
Belinfante first chose particularly simple paths at a fixed
Lorentz frame. They are straight line paths at constant time,
converging toward the field points $\bm{x}$ in arbitrary direction specified
by unit vector $\bm{\epsilon}$ :
\begin{equation}
 z^\nu (x, \xi) \ = \ x^\nu \ + \ \epsilon^\nu \,\xi ,
\end{equation}
where one can set without loss of generality as
\begin{equation}
 \epsilon^0 \ = \ 0, \ \ \ \bm{\epsilon}^2 \ = \ 1, \ \ \ 
 \xi \ = \ - \,r \ \equiv \ |\bm{z} - \bm{x}| .
\end{equation}
With this choice of path, Eq.(\ref{Eq:Atilde_Bel}) reduces to 
\begin{equation}
 \tilde{A}_\mu (x) \ = \ \int_{- \,\infty}^0 \,F_{i \mu} (z) \,\epsilon^i \,d \xi .
 \label{Eq:A_fixed_direction}
\end{equation}
This potential $\tilde{A}_\mu$ is by construction gauge-invariant but it depends on the
direction of the unit vector $\bm{\epsilon}$.
To remove this directional dependence, Belinfante propose to average Eq.(\ref{Eq:A_fixed_direction})
over all possible direction of $\bm{\epsilon}$ as
\begin{equation}
 \bar{A}_\mu (x) \ \equiv \ \int \,\tilde{A}_\mu (x) \,\frac{d \Omega}{4 \,\pi} ,
\end{equation}
where $d \Omega$ is the infinitesimal solid angle in the direction of $\bm{\epsilon}$.
By using the identities
\begin{equation}
 d^3 z \ = \ r^2 \,d r \,d \Omega ,
\end{equation}
and
\begin{equation}
 | d \xi | \ = \ | d r | \ = \ \frac{d^3 z}{r^2} \,d \Omega,
\end{equation}
it can be shown that $\bar{A}_\mu (x)$ reduces to the following form : 
\begin{equation}
 \bar{A}_\mu (x)  \ = \ 
 \int \,d^3 z \,F_{i \mu} (z) \,\frac{\partial}{\partial z_i} \,
 \left( \frac{1}{4 \,\pi \,r} \right) 
 \ = \ - \,\int \,d^3 z \,\left[ \nabla^i \,F_{i \mu} (z) \right] \,
 \left( \frac{1}{4 \,\pi \,r} \right) .
\end{equation}
If expressed with components, this reads as
\begin{equation}
 \bar{A}^0 (x) \ = \ \int d^3 z \,\frac{\nabla \cdot \bm{E} (\bm{z}, t)}{4 \,\pi \,r} , \ \ \ 
 \bar{\bm{A}} (x) \ = \ \int \,d^3 z \,\frac{\nabla \times \bm{B} (\bm{z}, t)}{4 \,\pi r} .
 \label{Eq:A0_Avec}
\end{equation}
One therefore finds that the resultant potential $\bar{A}_\mu (x)$ just coincides
with the vector potential in the Coulomb gauge. 
Thus, by averaging over the direction of straight-line paths, one could remove
the line-dependence from the original DeWitt gauge potential.
Besides, since it is expressed with the electric and magnetic fields only, it is obviously gauge-invariant. 
This observation lets us reconfirm the one and only
nature of the Coulomb gauge among plural gauge choices.
However, an attention must be paid to the the following facts. 
First, for that the above statement is meaningful, the integrals in
Eq.(\ref{Eq:A0_Avec}) must converge. The condition of convergence is
just the same as the condition of the Helmholtz theorem to hold.
Second, the above path averaging is carried out over the direction of the straight lines
at the {\it constant time} in a {\it fixed Lorentz frame}. 
This restriction is nothing serious
as far as we are dealing with the electrodynamics of non-relativistic charged particles
in a prescribed laboratory frame. This is the reason why the Coulomb gauge is the
most convenient choice for such problems. 
As discussed in \cite{Waka2015}, however, the situation
changes drastically for fully relativistic problems like the nucleon spin decomposition
problem.

In the nucleon spin decomposition problem, 
two new features enter into the game. 
First, the nucleon is a strongly-coupled bound state
of quarks and gluons described by a nonabelian gauge theory called
the quantum chromodynamics. 
Second, the only way to empirically verify the proposed nucleon
spin decomposition is to use deep-inelastic-scattering (DIS)
measurements, in which the relativity plays crucial roles.  
As to the first point, it is known that the idea of gauge-invariant
(but path-dependent) electron and photon field can be extended
to the nonabelian gauge theory with a slight
modification \cite{Ivanov-Korchemsky1985},\cite{Ivanov1986}. 
(The application of the gauge-invariant formulation of the nonabelian
gauge theory to the nucleon spin decomposition problem was
discussed in \cite{Waka2013},\cite{Lorce2013A},\cite{Lorce2013B}.)
More important here is the second point, i.e.  
the proper treatment of the relativistic kinematics \cite{Waka2015}.
Within the relativistic framework, the path for defining the
gauge-invariant gluon and quark fields can generally be chosen in
4-dimensional space-time manifold.
Here, instead of discussing the general path choices in the
gauge-invariant formulation of
nonabelian gauge theory, we confine to several simple but important
choices of path in the Minkowski space, which is widely believed to be
equivalent to working in the so-called physical gauges. 

The most popular physical gauges are the temporal gauge
($A^0 = 0$), the light-cone gauge ($A^+ = 0$ with $A^+ \equiv (A^0 + A^3) \,/\, \sqrt{2}$),
the spatial axial gauge ($A^3 = 0$), and the Coulomb
gauge ($\nabla \cdot \bm{A} = 0$).
We first recall the fact that selecting the physical component of the gauge field is
intimately connected with choosing a particular gauge.
In the general axial gauges, the physical component (or the gauge-covariant part)
of the gluon field is known to be expressed as
\begin{equation}
 A^\mu_{phys} (x) \ = \ n_\nu \,\int_{- \,\infty}^\infty \, d \lambda \,
 \left(\mp \,\theta (\pm \,\lambda)  \right) \,
 {\cal L} [x \,;\, x + n \,\lambda] \,F^{\mu \nu} (x + n \,\lambda) \,
 {\cal L} [ x + n \,\lambda \,;\, x] . \label{Eq:A_phys}
\end{equation}
Here, $F^{\mu \nu}$ is the field strength tensor of the gluon field,
while 
\begin{equation}
 {\cal L} [\,y \,;\, x] \ = \ {\cal P} \,\exp \,\left( i \,g \,
 \int_x^y \,d z_\mu \,A^\mu (z) \right) ,
\end{equation}
is a gauge-link (also called the
Wilson line) connecting the two space-time points $x$ and $y$ with
a straight line, ${\cal P}$ is the path-ordering operator, whereas
$n^\mu$ is a constant four-vector characterizing the direction of the path.
The dependence on the direction of $n^\mu$ is generally believed to reflect
the gauge-dependence of the idea of the physical component of the gauge field.
The familiar three physical gauges correspond to the following choice of $n^\mu$ : 
\begin{equation}
 n^\mu \ = \ \left\{ \begin{array}{lcl}
 \ \ (1, 0, 0, 0) \ & \ \Leftrightarrow \ & \ \mbox{temporal gauge}, \\
 \ \frac{1}{\sqrt{2}} \,(1, 0, 0, -1) \ & \ \Leftrightarrow \ & \ 
 \mbox{light-cone gauge}, \\
 \ \ (0, 0, 0, 1) \ & \ \Leftrightarrow \ & \ \mbox{spatial axial gauge}. \\
 \end{array} \right.
\end{equation}
In fact, with these choices of $n^\mu$, the physical component
of the gluon field defined by (\ref{Eq:A_phys}) satisfies the constraint,
\begin{equation}
 n_\mu \,A^\mu_{phys} \ = \ 0 ,
\end{equation}
which is nothing but the gauge-fixing conditions of the general
axial gauges.
Since there is no particular spatial direction in the Coulomb gauge-fixing condition,
the corresponding physical component may be obtained by averaging over
path-dependent potential over the direction of all straight line paths at constant
time, as in the abelian case.

Confining here to the three straight-line paths corresponding to the general
axial gauges, it can be readily verified that any choice of $n^\mu$ gives
the physical component of the gluon field, which transforms 
gauge-covariantly, i.e. as
\begin{equation}
 A^\mu_{phys} (x) \ \rightarrow \ U(x) \,A^\mu_{phys} (x) \,U^{- \,1} (x) ,
\end{equation}
under an arbitrary gauge transformations $U(x) = e^{\,i \,\omega_a (x) \,T_a}$.
This enables us to write down gauge-invariant expression of the gluon
spin as well as that of the g.i.c. OAM of quarks in the 
nucleon \cite{Hatta2011},\cite{Zhang-Pak2012}.
(More precisely, the gluon spin should be called the gluon helicity
or the longitudinal component of the gluon spin with respect to
the direction of the nucleon momentum.) 
They are given as
\begin{equation}
 \Delta G \ = \ \frac{1}{2 \,p^+} \,\langle p, s \,|\,
 2 \,\mbox{Tr} \,\left[ \epsilon^{j k}_\perp \,F^{j +} (0) \,A^k_{phys} (0) \right] \,
 |\,p, s \rangle ,
\end{equation}
and
\begin{equation}
 L^{g.i.c.}_z \ = \ \frac{{\cal N}^\prime}{2 \,p^+} \,
 \langle p, s \,|\, \int \,d^2 r_\perp \,\bar{\psi} (0, \bm{r}_\perp) \,\gamma^+ \,
 \frac{1}{i} \,\left( \bm{r}_\perp \times \bm{D}_{pure} \right)_z \,
 \psi (0, \bm{r}_\perp) \,|\, p, s \rangle ,
\end{equation}
where ${\cal N}^\prime = 1 \,/\,\left( \int \,d^2 r_\perp \right)$,
$\epsilon^{j k}_\perp \,(j, k = 1,2)$ is the antisymmetric tensor
in two spatial dimension, and $|\, p, s \rangle$ is the nucleon state
with momentum $p$ and spin $s$.
The pure-gauge covariant derivative is defined by
$\bm{D}_{pure} \equiv \nabla - i \,g \,\bm{A}_{pure}$ with
$\bm{A}_{pure} \equiv \bm{A} - \bm{A}_{phys}$.
Although the above quantities are formally gauge-invariant, they
are path-dependent, just because they contains the quantity
$\bm{A}_{phys}$, whose definition is path-dependent and non-local.
It should be contrasted with the expression of the mechanical
OAM of quarks given as
\begin{equation}
 L^{g.i.c.}_z \ = \ \frac{{\cal N}}{2 \,p^+} \,
 \langle p, s \,|\, \int \,d^3 r \,\bar{\psi} (\bm{r}) \,\gamma^+ \,
 \frac{1}{i} \,\left( \bm{r} \times \bm{D} \right)_z \,
 \psi (\bm{r}) \,|\, p, s \rangle ,
\end{equation}
where ${\cal N} = 1 \,/\,\left( \int \,d^3 r \right)$, while 
$\bm{D} = \nabla - i \,g \,\bm{A}$ is the standard
covariant derivative. Since the standard covariant derivative
contains the full gauge field, the definition of the
mechanical OAM is gauge-invariant as well as path-independent.
Also interesting to remember is the relation between the
g.i.c. OAM and the mechanical OAM. Using the definition
$\bm{A}_{pure} = \bm{A} - \bm{A}_{phys}$, one can readily
verify the following relation
\begin{equation}
 L^{g.i.c.}_z \ = \ L^{mech}_z \ + \ L^{pot}_z ,
\end{equation}
where
\begin{equation}
 L^{pot}_z \ = \ \frac{{\cal N}^\prime}{2 \,p^+} \,\langle p, s \,|\,
 \int \,d^2 r_\perp \,\bar{\psi} (0, \bm{r}_\perp) \,\gamma^+ \,g \,
 \left( \bm{r}_\perp \times \bm{A}_{phy} \right)_z \,\psi (0, \bm{r}_\perp)
 \,|\, p, s \rangle ,
\end{equation}
is nothing but the potential angular momentum in the terminology of
the papers \cite{Waka2010},\cite{Waka2011}.
One can then say that the path-dependence of the g.i.c. OAM
originates from that of the potential angular momentum,
which explicitly contains the path-dependent (or
direction-dependent) quantity $\bm{A}_{phys}$.

In any case, it would be true that the definition of the gluon spin as well as
that of the g.i.c. OAM are (at least formally) gauge-invariant
irrespectively of the choice of $n^\mu$.
This means that, from the viewpoint of gauge symmetry alone, these
three choices of $n^\mu$ is totally on an equal footing, and
anyone is neither superior nor inferior to the others.
However, an important lesson leaned from our analysis of the
Landau problem is a clear understanding about the meaning and the
consequence of the path choice. Namely, within 
the gauge-invariant but path-dependent
formulation of gauge theory {\it a la} DeWitt, what results from
a particular choice of path is a particular symmetry of the
quantum system or its wave function. 
We emphasize that this symmetry has
{\it nothing} to do with the gauge symmetry. 
In fact, remember that, in the Landau problem, the choice of the path $C_1$ in
the rectangular coordinate amounts to respecting the 
{\it translational symmetry} with respect to the $x$-axis. On the other hand, 
if we take the path $C_I$ in the cylindrical
coordinate, it amounts to respecting the axial symmetry
or the {\it rotational symmetry} around the $z$-axis.

The understanding above obtained from the Landau problem
gives a deep insight into our discussion of the nucleon
spin decomposition problem. The question here is what symmetry
is respected by a particular choice of the path in the 4-dimensional
Minkowski space. Undoubtedly, if we choose the straight-line
path along the $z$-direction, what is respected is the
rotational symmetry around the $z$-axis. On the other hand, if we
choose the straight-line path along the light-cone direction, an additional
symmetry is respected besides the above two-dimensional rotation
symmetry. It is the Lorentz boost-invariance with respect
to the $z$-axis, which is the momentum direction of the
parent nucleon. (For the detail, see \cite{Waka2015}.)
What symmetry is respected if we choose the straight-line path
along the time direction, then ?
Since there is no particular spatial direction for this choice,
this choice of path respects the full rotational symmetry in
the 3-dimensional space. (This is also expected from the known
intimate connection between the temporal gauge and the Coulomb
gauge. Remember that, within the framework of the gauge-invariant
formulation of DeWitt, the Coulomb gauge is obtained by averaging
over the directions of straight-line paths in constant-time plane,
so that there is no particular direction in three-dimensional space.)

Why is the boost-invariance along the $z$-direction so important ?
The reason is because the DIS measurement is the only way to
experimentally verify the proposed nucleon spin decomposition, and that
the DIS amplitude has a light-cone dominance \cite{Waka2015}.
In the inclusive or semi-inclusive DIS processes, the ejected parton 
(a quark or a gluon) after being struck by high-energy virtual photon
travels along the light cone with the speed of light.
The light-like gauge link with the choice 
$n^\mu = \frac{1}{\sqrt{2}} \,(1, 0, 0, - \,1)$, which is also
contained in the definition of $A^\mu_{phys}$, is known to
simulate the interaction between the ejected parton and the
residual target in the DIS processes. 
Very importantly, the gluon spin (or the g.i.c OAM of quarks) defined by
the physical component of the gluon with the light-like gauge link has
an important property, which must be satisfied by any
PDFs \cite{Collins_Book}. 
That is the Lorentz-boost invariance along the direction
of the parent nucleon. Putting it in another way,
the physics favors the path in the light-cone direction
in our nucleon spin decomposition problem \cite {Hatta2011}.

As a last comment, since there is a definite physical meaning in
the choice of path or the direction for defining the physical component
of the gluon field, not only the mechanical OAM but also the canonical
OAM are thought to have some definite physical
meaning \cite{Waka2015},\cite{Burkardt2013}. 
What these two OAMs really 
mean in the nucleon spin decomposition problem was already discussed
in a recent paper \cite{Waka2016}. It was shown there that what represents
the OAM of quarks inside the nucleon is the mechanical OAM not
the canonical OAM. The latter represents the OAM of a quark in the
asymptotic distance (well outside the nucleon) after leaving the
target nucleon in the semi-inclusive DIS scattering processes.
In this sense, it would be legitimate to say that the canonical type decomposition
of the nucleon spin, also called the Jaffe-Manohar decomposition of the
nucleon spin \cite{JM1990},\cite{BJ1998}, is not such a decomposition that
genuinely represents the intrinsic (or static) property of the nucleon .

\section{Summary and concluding remarks}
\label{Section:s6}

The delicacy of the gauge choice in the Landau problem partially stems
from a singular nature of the magnetic field, which uniformly spread over
the whole plane. In fact, if the magnetic field is confined in a finite domain,
the vector potential, or more precisely its transverse component,
is essentially uniquely fixed by the Helmholtz
theorem. As was explicitly shown in the case of infinitely long solenoid
with finite cross-sectional area, what emerges here is just
the gauge potential of symmetric gauge or the Coulomb gauge.
(To be more precise, in the two dimensional problem, 
there still remains a freedom to choose
the Bawin-Burnel type gauge, although it is a singular and
multi-valued gauge.)
The uniform magnetic field in the Landau problem allows entirely 
different gauge from the symmetric gauge, namely, the two types of
Landau gauge. 
The choice of the Landau gauges favors the rectangular coordinate
treatment of the problem, while the choice of the symmetric gauge does
the cylindrical coordinate treatment, so that they lead to the
eigen-functions with totally different appearances.
It turns out that, because of the degeneracy of the Landau levels in
both gauges, the eigen-functions in both gauges are not connected
by a simple gauge transformation. Rather, what are related to the
eigen-functions in the symmetric gauge by the gauge transformation is
a particular linear combination of the eigen-functions in the Landau gauge. 
We showed that the weight function of this linear combination is nothing
but the matrix element of the (unitary) gauge transformation operator
sandwiched with the eigen-states of the Landau gauge and those of
the symmetric gauge. We also derived the explicit form of this weight
function.

In addition to the above-mentioned standard formulation of the Landau
problem, we have also investigated the same problem on the basis of the
gauge-invariant but path-dependent formulation of the quantum
electrodynamics {\it a la} DeWitt.
A great advantage of this gauge-invariant formulation is that, with the
help of the path-dependent (non-local) phase factor, we can
solve the Schr\"{o}dinger equation without specifying explicit form of
the gauge potential. With the choice of the two polygonal line paths
in the rectangular coordinate, we are led to the eigen-functions expressed
by the Hermite polynomials except for a nonlocal phase factor.
It was shown that these choices of path have a close connection with the two types
of Landau gauge. On the other hand, with the choice of straight-line path
in the cylinder coordinate, we are led to the eigen-functions expressed
by the associated Laguerre polynomial, again except for nonlocal phase
factors. This choice of path then has an intimate connection with symmetric
gauge choice. Still, a highly nontrivial fact is that these latter eigen-functions
are given for arbitrary form of the vector potential, which especially
allows us to take the vector potential in either of the two Landau gauges.
In this Landau gauge limit, however, the above eigen-functions do not
reduce to the standard eigen-functions in the Landau gauge.
Rather, they reduce to a particular linear combination of the
eigen-functions in the Landau gauge, which is the same object
already obtained in our discussion of the standard (gauge-fixed) formulation of
the Landau problem. 

Through the study of DeWitt's gauge-invariant formulation of the
Landau problem,
we realize that the choice of path has an intimate connection with the
choice of gauge. For instance, the choice of a polygonal line path $C_1$ in the
rectangular coordinate system is shown to have inseparable connection with
the choice of the 1st Landau gauge, whereas the choice of the path $C_I$ in
the cylindrical coordinate has a close connection with the choice of symmetric
gauge. Nonetheless, highly nontrivial fact is that the choice of path is not
absolutely equivalent to the choice of gauge, because we still have freedom
to take any gauges for each choice of path. 
A question is therefore what is meant by the path choice in the
gauge-invariant but path-dependent formulation of DeWitt.
We conclude that the choice of path in DeWitt's formalism is equivalent to selecting 
a particular symmetry in our quantum mechanical problem.
As we have shown, the choice of the polygonal line path $C_1$ amounts to
respecting the translational symmetry with respect to the $x$-axis.
Once the gauge-invariant wave functions are constructed so as to satisfy
this symmetry, the expectation value of the g.i.c. momentum operator $K_x$
is shown to be independent of the three choices of gauges, i.e. the two
Landaus and the symmetric gauge at variance with that of the ordinary
canonical momentum operator $p_x$. Nevertheless, this gauge-independent
nature of the g.i.c. momentum should be taken with care. In fact, we compare
the expectation value of the g.i.c. momentum and that of the mechanical
momentum, thereby confirming the physical nature of the latter
as compared with the former.
On the other hand, if we choose the path $C_I$ in the cylindrical coordinate,
it amounts to respecting the rotational symmetry around the $z$-axis.
Once the gauge-invariant wave functions are constructed so as to meet this
symmetry, we find that the expectation value the g.i.c. OAM and that of
the standard canonical OAM are both independent of the three choices of
gauge and they just coincide. 
This is thought to be a consequence of the rotational symmetry
around the $z$-axis. However, the expectation value of these two types
of canonical OAM is different from that of the mechanical OAM. 
The physical nature of the mechanical OAM (as compared with the others)
is thought to manifest in its relation with the observable Landau energies.

We have also discussed the issue of gauge choice in the nucleon spin
decomposition problem from the general viewpoint of the gauge-invariant
but path-dependent formulation of nonabelian gauge theory.
As we have argued, for relativistic problems like the nucleon spin
decomposition problem, the path for defining the {\it physical component}
of the gauge field can be chosen arbitrarily in the 4-dimensional
Minkowski space. 
Among several simple choices of paths, i.e. the straight line
in the time direction, the light-cone direction, and the $z$-direction,
only the light-cone path choice can satisfy the required symmetries
of the physics in question, i.e. the rotational symmetry
around the $z$-axis as well as the Lorentz-boost invariance along the same
direction. Accordingly, only the gluon spin operator (or g.i.c OAM of quarks)
defined by the physical component of the gluon with the light-like gauge link 
can satisfy the Lorentz-boost invariance along the momentum direction
of the parent nucleon, which must be respected by any parton
distribution functions. In other words, what selects the proper
definition of the physical component of the gluon field in the
nucleon spin decomposition problem is the {\it Lorentz symmetry} rather
than the {\it gauge symmetry} \cite{Waka2015}. We recall that the importance
of Lorentz symmetry in the proper definition of the physical component of the
gluon field was already stressed by Tiwari at the early stage
of debate \cite{Tiwari2008}.

\noindent
\section*{Acknowledgement}

\vspace{2mm}
\noindent
M.W. would like to thank S.C.~Tiwari for several useful comments.
M.W. also thanks the Institute of Modern Physics of the Chinese
Academy of Sciences in Lanzhou for hospitality.
Y.K. and P.-M.Z. are supported by the National Natural
Science Foundation of China (Grant No.11575254).
This work is partly supported by the Chinese Academy of Sciences
President's International Fellowship Initiative 
(No. 2018VMA0030 and No. 2018PM0028)



\appendix

\section{Calculation of gauge transformation matrix}

In this appendix, we evaluate the matrix element of the gauge transformation
operator between the eigenstates in the Landau gauge and those in the
symmetric gauge.
Here, we consider the $m \geq 0$ case, since the $m < 0$ case can be handled
similarly,
\begin{eqnarray}
 \langle \Psi^{(L)}_{n^\prime, k_y} \,| \,e^{\,- \,i \,\frac{1}{2} \,e \,B \,x \,y} \,| \,
 \Psi^{(S)}_{n,m} \rangle 
 \!\! &=& \!\! \int \,d x \,d y \,\,\frac{1}{\sqrt{2 \,\pi}} \,e^{\,- \,i \,k_y \,y} \,
 N_{n^\prime} \,H_{n^\prime} \left( \frac{x - x_0}{l_B} \right) \,
 e^{\,- \,\frac{(x - x_0)^2}{2 \,l_B^2}} \nonumber \\
 \times \ e^{\,- \,i \,\frac{1}{2} \,e \,B \,x \,y} \,\,
 \frac{1}{\sqrt{2 \,\pi}} \!\!\!\!\!\! &\,& \!\!\!\!\!\! N_{n,m} \,e^{\,i \,m \,\phi} \,
 \left( \frac{r^2}{2 \,l_B^2} \right)^{m/2} \,
 e^{\,- \,\frac{r^2}{4 \,l_B^2}} \,L^m_{n-m} \left( \frac{r^2}{2 \,l_B^2} \right).
 \hspace{5mm}
\end{eqnarray}
Using the notation $z = x + i \,y = r \,e^{\,i \,\phi}$ and $\bar{z} = x - i \,y$,
we rewrite it as
\begin{eqnarray}
 \langle \Psi^{(L)}_{n^\prime, k_y} \,| \,e^{\,- \,i \,\frac{1}{2} \,e \,B \,x \,y} \,| \,
 \Psi^{(S)}_{n,m} \rangle 
 &=& \frac{1}{2 \,\pi} \,\int \,d x \,d y \,\,e^{\,- \,i \,k_y \,y} \,
 N_{n^\prime} \,H_{n^\prime} \left( \frac{x - x_0}{l_B} \right) \,
 e^{\,- \,\frac{(x - x_0)^2}{2 \,l_B^2}} \nonumber \\
 \times \ e^{\,- \,i \,\frac{1}{2} \,e \,B \,x \,y}
 &\times& N_{n,m} \,\,2^{m/2} \,\left( \frac{z}{2 \,l_B} \right)^m \,
 e^{\,- \,\frac{\bar{z} \,z}{4 \,l_B^2}} \,L^m_{n-m} \,
 \left( \frac{\bar{z} \,z}{2 \,l_B^2} \right).
 \hspace{5mm}
\end{eqnarray}
Now, with use of the mathematical identity
\begin{equation}
 \int_{- \,\infty}^\infty \,d u \,e^{- \,u^2} \,H_p (u + v) \,H_q (u + w) \ = \ 
 \sqrt{\pi} \,2^q \,p ! \,w^{q - p} \,L^{q - p}_p (- \,2 \,v \,w) \ \ \ \ 
 (\mbox{for} \ \ p \leq q), \label{Eq:HL_identity}
\end{equation}
we can write as
\begin{eqnarray}
 &\,& \!\!\! \left( \frac{z}{2 \,l_B} \right)^m \,L^m_{n - m} \left( \frac{\bar{z} \,z}{2 \,l_B^2} \right)
 \ = \ \!\! \frac{(-1)^m}{\sqrt{\pi} \,2^n \,(n - m) !} \,\,
 \int_{- \,\infty}^\infty \,d u \,e^{\,- \,u^2} \,H_{n-m} \left( u + \frac{\bar{z}}{2 \,l_B} \right)
 \,H_n \left( u - \frac{z}{2 \,l_B} \right). \hspace{10mm} 
\end{eqnarray}
Thus we find that
\begin{eqnarray}
 \langle \Psi^{(L)}_{n^\prime, k_y} \,| \,e^{\,- \,i \,\frac{1}{2} \,e \,B \,x \,y} \,| \,
 \Psi^{(S)}_{n,m} \rangle 
 \!\! &=& \!\! \frac{1}{2 \,\pi} \,\int \,d x \,d y \,\,e^{\,- \,i \,k_y \,y} \,
 N_{n^\prime} \,H_{n^\prime} \left( \frac{x - x_0}{l_B} \right) \,
 e^{\,- \,\frac{(x - x_0)^2}{2 \,l_B^2}} \nonumber \\
 \!\! &\times& \!\! e^{\,- \,i \,\frac{1}{2} \,e \,B \,x \,y}
 \times \,N_{n,m} \,\,2^{m/2} \,\frac{(-1)^m}{\sqrt{\pi} \,2^n \,(n - m) !} \,
 e^{\,- \,\frac{\bar{z} \,z}{4 \,l_B^2}} \nonumber \\
 \!\! &\times& \!\!\!\!  
 \int_{- \,\infty}^\infty \,d u \,e^{\,- \,u^2} \,H_{n-m} \left( u + \frac{\bar{z}}{2 \,l_B} \right)
 \,H_n \left( u - \frac{z}{2 \,l_B} \right). \hspace{4mm}
 \hspace{8mm}
\end{eqnarray}
To proceed, first we note that
\begin{equation}
 e^{\,- \,i \,k_y \,y} \,e^{\,- \,i \,\frac{1}{2} \,e \,B \,x \,y} \ = \ 
 e^{\,- \,i \,\frac{1}{2 \,l_B^2} \,(x - 2 \,x_0) \,y} . 
\end{equation}
Next, we introduce a variable transformation from $u$ to $w = u - i \,y \,/\,(2 \,l_B)$,
which gives
\begin{eqnarray}
 u \ + \ \frac{\bar{z}}{2 \,l_B} \ = \ u \ - \ \frac{i \,y}{2 \,l_B} \ + \ \frac{x}{2 \,l_B}
 \ = \ w \ + \ \frac{x}{2 \,l_B} , \\
 u \ - \ \frac{z}{2 \,l_B} \ = \ u \ - \ \frac{i \,y}{2 \,l_B} \ - \ \frac{x}{2 \,l_B}
 \ = \ w \ - \ \frac{x}{2 \,l_B} .
\end{eqnarray}
Now, using the manipulations
\begin{eqnarray}
 &\,& e^{\,- \,u^2} \ = \ e^{\,- \,\left( w + \frac{i \,y}{2 \,l_B} \right)^2} \ = \ 
 e^{\,- \,w^2} \,e^{\,- \,i \,\frac{w \,y}{l_B}} \,e^{\,\frac{y^2}{4 \,l_B^2}} , \\
 &\,& e^{\,- \,\frac{(x - x_0)^2}{2 \,l_B^2}} \,e^{\,- \,\frac{\bar{z} \,z}{4 \,l_B^2}} \,e^{\,- \,u^2}
 \ = \ e^{\,- \,\frac{(x - x_0)^2}{2 \,l_B^2}} \,e^{\,- \,\frac{x^2}{4 \,l_B^2}} \,
 e^{\,- \,w^2} \,e^{\,- \,i \,\frac{w \,y}{l_B}} ,
\end{eqnarray}
as well as
\begin{equation}
 e^{\,- \,i \,\frac{1}{2 \,l_B^2} \,(x - 2 \,x_0) \,y} \,e^{\,- \,i \,\frac{w \,y}{l_B}} \ = \ 
 e^{\,- \,i \,\frac{1}{l_B} \,\left( w \,+ \,\frac{x - 2 \,x_0}{2 \,l_B} \right) \,y} ,
\end{equation}
we obtain
\begin{eqnarray}
 \langle \Psi^{(L)}_{n^\prime, k_y} \,| \,e^{\,- \,\frac{1}{2} \,i \,e \,B \,x \,y} \,| \,
 \Psi^{(S)}_{n,m} \rangle 
 &=& \frac{1}{2 \,\pi} \,N_{n^\prime} \,N_{n,m} \,\frac{(-1)^m \,2^{m/2}}{\sqrt{\pi} \,2^n \,(n - m) !} \,
 \int \,d x \,d y \,\, \int \,d w \,H_{n^\prime} \left( \frac{x - x_0}{l_B} \right) \nonumber \\
 &\times& e^{\,- \,i \,\frac{1}{l_B} \,\left( w \,+ \,\frac{x - 2 \,x_0}{2 \,l_B} \right) \,y} \,
 e^{\,- \,\frac{(x - x_0)^2}{2 \,l_B^2}} \,e^{\,- \,\frac{x^2}{4 \,l_B^2}} \,e^{\,- \,w^2} \nonumber \\
 &\times& H_{n-m} \left( u + \frac{\bar{z}}{2 \,l_B} \right) \,\,H_n \left( u - \frac{z}{2 \,l_B} \right) .
 \hspace{5mm}
\end{eqnarray}
The integration over the variable $w$ can readily be done as
\begin{equation}
 \int_{- \,\infty}^\infty \,d y \,e^{\,- \,i \,\frac{1}{l_B} \,
 \left( w + \frac{x - 2 \,x_0}{2 \,l_B} \right) \,y} \ = \ 2 \,\pi \,l_B \,
 \delta \left( w + \frac{x - 2 \,x_0}{2 \,l_B} \right) ,
\end{equation}
which makes the integration over $w$ trivial, thereby leading to the result : 
\begin{eqnarray}
 \langle \Psi^{(L)}_{n^\prime, k_y} \,| \,e^{\,- \,\frac{1}{2} \,i \,e \,B \,x \,y} \,| \,
 \Psi^{(S)}_{n,m} \rangle 
 \!\! &=& \!\! \frac{1}{2 \,\pi} \,N_{n^\prime} \,N_{n,m} \,l_B \,
 \frac{(-1)^m \,2^{m/2}}{\sqrt{\pi} \,2^n \,(n - m) !} \,
 \int \,d x \,H_{n^\prime} \left( \frac{x - x_0}{l_B} \right) \nonumber \\
 \!\! &\times& \!\! e^{\,- \,\frac{(x - x_0)^2}{2 \,l_B^2}} \,e^{\,- \,\frac{x^2}{4 \,l_B^2}} \,
 e^{\,- \,\frac{(x - 2 \,x_0)^2}{4 \,l_B^2}} \,
 H_{n-m} \left( \frac{x_0}{l_B} \right) \,H_n \left( - \, \frac{x - x_0}{l_B} \right).
 \hspace{12mm}
\end{eqnarray}
Using the relation
\begin{equation}
 e^{\,- \,\frac{(x - x_0)^2}{2 \,l_B^2}} \,e^{\,- \,\frac{x^2}{4 \,l_B^2}} \,
 e^{\,- \,\frac{(x - 2 \,x_0)^2}{4 \,l_B^2}} \ = \ 
 e^{\,- \,\frac{(x - x_0)^2}{l_B^2}} \,e^{\,- \,\frac{x_0^2}{2 \,l_B^2}} ,
\end{equation}
and the identity $H_n (- \,x) = (- \,1)^n \,H_n (x)$ of the Hermite polynomial, 
we therefore get
\begin{eqnarray}
 \langle \Psi^{(L)}_{n^\prime, k_y} \,| \,e^{\,- \,i \,\frac{1}{2} \,e \,B \,x \,y} \,| \,
 \Psi^{(S)}_{n,m} \rangle 
 \!\! &=& \!\! \frac{1}{2 \,\pi} \,N_{n^\prime} \,N_{n,m} \,l_B \,
 \frac{(-1)^{n+m} \,2^{m/2}}{\sqrt{\pi} \,2^n \,(n - m) !} \,
 H_{n-m} \left( \frac{x_0}{l_B} \right) \,e^{\,- \,\frac{x_0^2}{2 \,l_B^2}} \nonumber \\
 \!\! &\times& \!\! \int \,d x \,e^{\,- \,\frac{(x - x_0)^2}{l_B^2}} \,
 \,H_{n^\prime} \left( \frac{x - x_0}{l_B} \right) \,H_n \left( \frac{x - x_0}{l_B} \right).
 \hspace{12mm}
\end{eqnarray}
Now, by using the familiar ortho-normalization relation of the Hermite polynomial
as well as the definitions of the normalization constants $N_n$ and $N_{n,m}$, we
arrive at the desired relation,
\begin{equation}
 \langle \Psi^{(L)}_{n^\prime, k_y} \,| \,e^{\,- \,i \,\frac{1}{2} \,e \,B \,x \,y} \,| \,
 \Psi^{(S)}_{n,m} \rangle 
 \ = \ \delta_{n^\prime, n} \,U_{n,m} (x_0),
\end{equation}
where
\begin{equation}
 U_{n,m} (x_0) \ = \ C_{n,m} \,H_{n-m} \left( \frac{x_0}{l_B} \right) \,
 e^{\,- \,\frac{x_0^2}{2 \,l_B^2}} ,
\end{equation}
with
\begin{equation}
 C_{n,m} \ = \ l_B \,\left( \frac{1}{\sqrt{\pi} \,2^{n-m} \,(n - m) ! \,l_B} \right)^{1/2} .
\end{equation}
%

\noindent
\section{Evaluation of the weighted integral}

Here we show the calculation of the weighted integral ${\cal I}$ in Eq.(\ref{Eq:U-Unm}),
\begin{equation}
 {\cal I} \ = \ \int \,d k_y \,U_{n,m} (x_0) \,\Psi^{(L)}_{n,k_y} (x,y).
\end{equation}
We again assume $m \geq 0$, since the $m < 0$ case can be treated entirely
in the same way.
Inserting the expressions of $U_{m,n} (x_0)$ and $\Psi^{(L)}_{n,k_y} (x,y)$ into
the above equation, and using the relation $x_0 = - \,l_B^2 \,k_y$, we have
\begin{eqnarray}
 {\cal I} &=& \frac{1}{l_B^2} \,C_{n,m} \,\int \,d x_0 \,
 H_{n-m} \left( \frac{x_0}{l_B} \right) \,e^{- \,\frac{x_0^2}{2 \,l_B^2}}
 \, \, \frac{1}{\sqrt{2 \,\pi}} \,N_n \,e^{\,i \,k_y \,y} \,
 H_{n} \left( \frac{x -x_0}{l_B} \right) \,e^{- \,\frac{(x - x_0)^2}{2 \,l_B^2}} .
\end{eqnarray}
Using the equality
\begin{equation}
 e^{- \,\frac{x_0^2}{2 \,l_B^2}} \,e^{\,i \,k_y \,y} \,e^{\,- \,\frac{(x - x_0)^2}{2 \,l_B^2}}
 \ = \ e^{- \,\frac{1}{l_B^2} \,\left( x_0 - \frac{\bar{z}}{2} \right)^2} \,
 e^{- \,\frac{1}{4 \,l_B^2} \,\bar{z} \,z} \,e^{- \,i \,\frac{1}{2} \,e \,B \,x \,y} ,
\end{equation}
we thus obtain
\begin{eqnarray}
 {\cal I} &=& \frac{1}{l_B^2} \,C_{n,m} \,\frac{1}{\sqrt{2 \,\pi}} \,N_n \,
 e^{- \,\frac{1}{4 \,l_B^2} \,\bar{z} \,z} \,e^{- \,i \,\frac{1}{2} \,e \,B \,x \,y} \nonumber \\
 &\,& \times \ \int \,d x_0 \,e^{- \,\frac{1}{l_B^2} \,\left( x_0 - \frac{\bar{z}}{2} \right)^2} \,
 H_{n-m} \left( \frac{x_0}{l_B} \right) \,H_{n} \left( \frac{x - x_0}{l_B} \right) .
\end{eqnarray}
After the variable transformation from $x_0$ to $u$ with
\begin{equation}
 x_0 \ - \ \frac{\bar{z}}{2} \ = \ l_B \,u,
\end{equation}
the above expression can be transformed into the form,
\begin{eqnarray}
 {\cal I} &=& \frac{1}{l_B} \,(- \,1)^n \,C_{n,m} \,\frac{1}{\sqrt{2 \,\pi}} \,N_n \,
 e^{- \,\frac{1}{4 \,l_B^2} \,\bar{z} \,z} \,e^{- \,i \,\frac{1}{2} \,e \,B \,x \,y} \nonumber \\
 &\,& \times \ \int \,d u \,e^{\,- \,u^2} \,H_{n-m} \left( u + \frac{\bar{z}}{2 \,l_B} \right) \,
 H_n \left( u - \frac{z}{2 \,l_B} \right) .
\end{eqnarray}
Using the identity (\ref{Eq:HL_identity}), we therefore get
\begin{eqnarray}
 &\,& \int \,d u \,e^{\,- \,u^2} \,H_{n-m} \left( u + \frac{\bar{z}}{2 \,l_B} \right) \,
 H_n \left( u - \frac{z}{2 \,l_B} \right) \nonumber \\
 &\,& \ = \ (- \,1)^m \,2^n \,\sqrt{\pi} \,(n - m) ! \,
 \left( \frac{z}{2 \,l_B} \right)^m \,L^m_{n-m} \left( \frac{\bar{z} \,z}{2 \,l_B^2} \right). 
\end{eqnarray}
This gives
\begin{eqnarray}
 {\cal I} &=& \frac{1}{l_B} \,(- \,1)^n \,C_{n,m} \,\frac{1}{\sqrt{2 \,\pi}} \,N_n \,
 e^{- \,\frac{1}{4 \,l_B^2} \,\bar{z} \,z} \,e^{- \,i \,\frac{1}{2} \,e \,B \,x \,y} \nonumber \\
 &\,& \times \ (- \,1)^m \,2^n \,\sqrt{\pi} \,(n - m) ! \,\left( \frac{z}{2 \,l_B} \right)^m \,
 L^m_{n-m} \left( \frac{\bar{z} \,z}{2 \,l_B^2} \right) .
\end{eqnarray}
Finally, with use of the definitions of the coefficients $C_{n,m}$ and $N_n$, we find that
\begin{equation}
 {\cal I} \ \equiv \ \int \,d k_y \,U_{n,m} (x_0) \,\Psi^{(L)}_{n,k_y} (x,y) 
 \ = \ 
 e^{- \,i \,\frac{1}{2} \,e \,B \,x \, y}
 \,\Psi^{(S)}_{n,m} (x,y) .
\end{equation}

\noindent
\section{Proof of gauge-invariance of the electron wave function 
$\tilde{\Psi}^{(C_{\rm I})} (x,y)$ given by Eq.(\ref{Eq:Rep2})}

\begin{figure}[ht]
\begin{center}
\includegraphics[width=6cm]{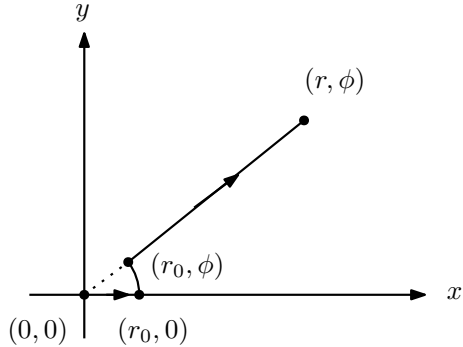}
\caption{The path $C_{\rm I}$ defined as $r_0 \rightarrow 0$ limit of the path
illustrated in this figure.}
\label{Fig:limiting_path}
\end{center}
\end{figure}

Here we prove that the electron wave function $\tilde{\Psi}^{(C_{\rm I})} (x,y)$
defined by
\begin{equation}
 \tilde{\Psi}^{(C_{\rm I})} (x,y) \ \equiv \ U_{I}^{- \,1} \,\Psi (x,y) ,
\end{equation}
with $U_{\rm I} \ = \ e^{\,- \,i \,e \,\int_0^r \,A_r (r^\prime, \phi) \,d r^\prime}$
is gauge-invariant. Under an arbitrary gauge transformation
\begin{eqnarray}
 \Psi (r, \phi) &\rightarrow& e^{\,- \,i \,e \,[\omega (r, \phi) - \omega (0,0)]}
 \,\Psi (r, \phi) ,\nonumber \\
 \bm{A} (r \,\phi) &\rightarrow& \bm{A} (r, \phi) \ + \ \nabla \omega (r, \phi) ,
\end{eqnarray}
the radial and azimuthal components of $\bm{A}$ transforms as
\begin{eqnarray}
 A_r (r, \phi) &\rightarrow& A_r (r, \phi) \ + \ 
 \frac{\partial \omega (r, \phi)}{\partial r} , \\
 A_\phi (r, \phi) &\rightarrow& A_\phi (r, \phi) \ + \ 
 \frac{1}{r} \,\frac{\partial \omega (r, \phi)}{\partial \phi} .
\end{eqnarray}
We first investigate the gauge-transformation property of the operator
$U_{\rm I}^{- \,1}$. Only one delicate point here is that
the line-integral contained in the definition of $U_{\rm I}^{- \,1}$ should be
interpreted as $r_0 \rightarrow 0$ limit of the following expression
(see Fig.\ref{Fig:limiting_path}) ; 
\begin{equation}
 U_{\rm I}^{- \,1} \ = \ \lim_{r_0 \rightarrow 0} \,
 e^{\, i \,e \,\left[ \int_{r_0}^r \,A_r (r^\prime, \phi) \,d r^\prime \ + \ 
 \int_0^\phi \,r_0 \,A_\phi (r_0, \phi^\prime) \,d \phi^\prime \right]} .
\end{equation}
%

Under the gauge transformation, the phase factor transforms as
\begin{eqnarray}
 &\,& \int_{r_0}^r \,A_r (r^\prime, \phi) \,d r^\prime \ + \ 
 \int_0^\phi \,r_0 \,A_\phi (r_0, \phi^\prime) \,d \phi^\prime \nonumber \\
 &\rightarrow&
 \int_{r_0}^r \,\left( A_r (r^\prime, \phi) \ + \ \frac{\partial}{\partial r^\prime} \,
 \omega (r^\prime, \phi) \right) \,d r\prime 
 \ + \ \int_0^\phi r_0 \,
 \left( A_\phi (r_0. \phi^\prime) \ + \ \frac{1}{r_0} \,
 \frac{\partial \omega (r_0, \phi^\prime)}{\partial \phi^\prime} \right)
 \,d \phi^\prime \hspace{10mm} \nonumber \\
 &=&  \int_{r_0}^r \,A_r (r^\prime, \phi) \,d r^\prime \ + \ 
 \int_0^\phi \,r_0 \,A_\phi (r_0, \phi^\prime) \,d \phi^\prime 
 \nonumber \\
 &\,& \hspace{20mm} \ + \ [\omega (r, \phi) \ - \ \omega(r_0, \phi)] 
 \ + \ 
 [\omega (r_0, \phi) \ + \ \omega (0,0)] \nonumber \\
 &=&  \int_{r_0}^r \,A_r (r^\prime, \phi) \,d r^\prime \ + \ 
 \int_0^\phi \,r_0 \,A_\phi (r_0, \phi^\prime) \,d \phi^\prime \ + \ 
 [\omega (r, \phi) \ - \ \omega (0,0) ] .
\end{eqnarray}
In the limit $r_0 \rightarrow 0$, this reduces to
\begin{equation}
 \int_0^r \,A_r (r^\prime, \phi) \,d r^\prime \ + \ 
 [\omega (r, \phi) \ - \ \omega (0, 0) ] .
\end{equation}
Thus, we find that, under the gauge transformation,
\begin{eqnarray}
 U_{\rm I}^{- \,1} &\rightarrow& U_{\rm I}^{- \,1} \,
 e^{\,i \,e \,[\omega (r, \phi) \ - \ \omega (0,0)]} , \\
 \Psi (r, \phi) &\rightarrow&
 e^{\,- \,i \,e \,[\omega (r, \phi) \ - \ \omega (0,0)]} \,
 \Psi (r, \phi) ,
\end{eqnarray}
and consequently
\begin{equation}
 \tilde{\Psi}^{(C_{\rm I})} (r,\phi) \ \equiv \ U_{\rm I}^{- \,1} \,\Psi (r,\phi)
 \ \rightarrow \ \tilde{\Psi}^{(C_{\rm I})} (r, \phi) ,
\end{equation}
which proves that $\tilde{\Psi}^{(C_{\rm I})} (r, \phi)$ is gauge-invariant.

\noindent
\section{The proof of the relation 
$\langle \Psi^{(C_I)}_{n,m} \,|\,
L^{can}_z \,|\,\Psi^{(C_I)}_{n,m} \rangle = m$ for arbitrary gauge.}

With use of the expression (138) for $\Psi^{(C_I)}_{n,m} (r, \theta)$,
we obtain
\begin{eqnarray}
 &\,& \langle \Psi^{(C_I)}_{n,m} \,|\,
 L^{can}_z \,|\,\Psi^{(C_I)}_{n,m} \rangle \ = \ 
 \langle \Psi^{(C_I)}_{n,m} \,|\,
 - \,i \,\frac{\partial}{\partial \phi} \,\,|\,\Psi^{(C_I)}_{n,m} \rangle \nonumber \\
 &=& \int_0^\infty \,d r \,r \,\int_0^{2 \,\pi} \,d \phi \,
 R_{n,m} (r) \,\left( m \ + \ \frac{1}{2} \,e \,B \,r^2 \ - \ 
 e \,r \,A_\phi (r, \phi) \right) \,R_{n,m} (r).
\end{eqnarray}
Here, we make use of the relation 
\begin{equation}
 r \,A_\phi (r,\phi) \ =  \frac{1}{2} \,r^2 \,B \ + \ \int_0^r \,d r^\prime \,
 \frac{\partial}{\partial \phi} \,A_r (r^\prime, \phi) ,
\end{equation}
which is obtained from the circular coordinate representation of the
identity $B = (\nabla \times \bm{A})_z$. This gives
\begin{equation}
 \langle \Psi^{(C_I)}_{n,m} \,|\,L^{can}_z \,|\,\Psi^{(C_I)}_{n,m} \rangle \ = \ 
 m \ - \ e \,\int_0^\infty \,d r \,r \,\int_0^{2 \,\pi} \,d \phi \,R_{n,m} (r) \, 
 \int_0^r \,d r^\prime \,\frac{\partial}{\partial \phi} \,A_r (r^\prime, \phi) \,
 R_{n,m} (r) .
\end{equation}
Carrying out the integral over $\phi$ first, we find that
\begin{equation}
 \int_0^{2 \,\pi} \,d \phi \,\frac{\partial}{\partial \phi} \,A_r (r^\prime, \phi)
 \ = \ A_r (r^\prime, 2 \,\pi) \ - \ A_r (r^\prime, 0).
\end{equation}
This vanishes, since it holds that
\begin{equation}
 A_r (r, \phi + 2 \,\pi) \ = \ A_r (r, \phi), \hspace{8mm}
 A_\phi (r, \phi + 2 \,\pi) \ = \ A_\phi (r, \phi) ,
\end{equation}
for arbitrary single-valued (or regular) gauge-field configuration. In this way, we obtain
\begin{equation}
 \langle \Psi^{(C_I)}_{n,m} \,|\,L^{can}_z \,|\,\Psi^{(C_I)}_{n,m} \rangle \ = \ m ,
\end{equation}
for any single-valued gauge.



\vspace{8mm}
\noindent
\section*{References}

\end{document}